\documentclass[useAMS,usenatbib]{mn2e}

\usepackage{amsmath,epsfig}

 \let\url\relax

\newcommand{\sinc}{{\rm sinc}}

\title{The GMRT EoR Experiment: Limits on Polarized Sky Brightness at 150 MHz}

\author[Pen et al.]{Ue-Li Pen$^1$, Tzu-Ching Chang$^{1,2}$,  
Christopher M. Hirata$^3$,
Jeffrey B. Peterson$^4$, 
\newauthor
Jayanta Roy$^5$, Yashwant Gupta$^5$, Julia Odegova$^1$, Kris Sigurdson$^6$
\\
$^{1}$ CITA, University of Toronto, 60 St.George St., Toronto, ON M5S 3H8, Canada \\
$^{2}$ Institute for Astronomy and Astrophysics, Academia Sinica, P.O. Box 23-141, Taipei 10617, Taiwan  \\
$^{3}$ M/C 130-33, California Institute for Technology, Pasadena, 
CA 91125, USA.\\
$^{4}$ Department of Physics, Carnegie Mellon University, 5000 Forbes Ave, Pittsburgh, PA 15213, USA \\
$^{5}$ National Center for Radio Astrophysics, Tata Institute for 
Fundamental Research, Pune 411 007, India \\
$^{6}$ Department of Physics and Astronomy, University of British Columbia, Vancouver, BC V6T 1Z1, Canada 
}

\begin{document}

\date{\today}

\maketitle

\begin{abstract}

The GMRT reionization effort aims to map out the large scale structure
of the Universe during the epoch of reionization (EoR).  Removal of
polarized Galactic emission is a difficult part of any 21 cm EoR program,
and we present new upper limits to diffuse polarized foregrounds at 150
MHz.  We find no high significance evidence of polarized emission
in our observed field at mid galactic latitude (J2000 08h26m+26).
We find an upper limit on the 2-dimensional
angular power spectrum of diffuse polarized foregrounds
of $[l^2C_l/2\pi]^{1/2}<3\,$K in frequency bins of width $\delta\nu=1\,$MHz at $300<l<1000$.
The 3-dimensional power spectrum of polarized emission, which is most directly
relevant to EoR observations, is
$[k^3P_P(k)/2\pi^2]^{1/2}<2\,$K
at $k_\perp>0.03h\,$Mpc$^{-1}$, $k<0.1h\,$Mpc$^{-1}$.  This can be compared to the expected
EoR signal in total intensity of $[k^3P(k)/2\pi^2]^{1/2}\sim 10\,$mK.
We find polarized
structure is substantially weaker than suggested by extrapolation from
higher frequency observations, so the new low upper limits reported here
reduce the anticipated impact of these foregrounds on EoR experiments.
We discuss  Faraday beam and depth depolarization models and compare
predictions of these models to our data. We report on a new technique
for polarization calibration using pulsars, as well as a new technique
to remove broadband radio frequency interference.  Our data indicate
that, on the edges of the main beam at GMRT, polarization squint creates
$\sim$ 3\% leakage of unpolarized power into polarized maps at zero
rotation measure. Ionospheric rotation was largely stable during these
solar minimum night time observations.

\end{abstract}



\section{Introduction}

A current frontier in observational and theoretical cosmology is the
search for and understanding of large scale structure during the epoch
of reionization (EoR).  Detection of reionization at $z\sim 10$ would allow study of 
the emergence of the first abundant
luminous objects. Such observations holds the promise of examination of 
astrophysical and cosmological processes
at epochs as early as a few hundred million years after the start of
the expansion of the Universe.

The WMAP satellite has measured polarization in the Cosmic Microwave
Background (CMB) at large angular scales.  This polarization is
believed to arise from Thomson scattering of the CMB photons near the
EoR \citep{2008arXiv0803.0593N,2008arXiv0803.0547K}. The observed
optical depth $\tau\sim 0.089\pm 0.016$ corresponds to an instantaneous
reionization redshift of $z_{\rm reion}=10.8\pm 1.4$.

One way to study the reionization transition in detail is by imaging redshifted
21cm emission. At redshifts above the EoR transition the gas is
neutral and is predicted to glow with about  a 25 mK sky brightness
temperature. After reionization is complete this glow is absent.
At redshifts close to the transition a patchy sky is expected.
Simulations \citep{2008MNRAS.tmp...77I} suggest that with existing
telescopes a measurement near 150 MHz may allow for statistical detection
of $\sim$ 20 Mpc patchiness in the neutral hydrogen.  This detection
would pin down the reionization redshift and begin the process of more
detailed study of the transition.

The 21 cm EoR signal, in the 10 milliKelvin range, is much smaller
than the expected $\sim 10$K patchiness due to Galactic synchrotron
emission and extragalactic foregrounds. One of the most difficult
challenges of 21 cm EoR astronomy is the removal of these bright
foregrounds.  The proposed strategy makes use of the very different
spatial and spectral character of the two components of sky
structure. Since the received frequency of the 21 cm line emission
encodes the distance of the source, the EoR signal consists of
a series of source screens that are largely independent. For example,
if one detects a patchy layer of cosmic material at 150 MHz due to the
EoR transition, another patchy layer must be present at 151 MHz, but
these two images come from distinct radial shells and the bright and
dim patches in the two images are not expected to be aligned on the
sky.  The characteristic size of the ionization structures near the epoch of
ionized-cell overlap is approximately 10 $h^{-1}$ Mpc, which corresponds to 5 arc
minutes in angular size, or about 1 MHz in radial distance for the
redshifted 21cm line at $z \sim 9$.  In contrast, the foreground
emission is highly correlated from one frequency to the next, since most
radio-bright structures emit with a very smooth synchrotron
spectrum. To remove the foregrounds some type of
frequency-differencing technique will be needed.

At frequencies above 300 MHz many radio sources with a synchrotron
spectrum have been shown to be linearly polarized, and this raises a
concern for the frequency differencing scheme.  In propagation through
the interstellar medium the polarization rotates on the sky because of
the Faraday effect.  In the polarization basis of the telescope this means
that each linear polarization component can oscillate with frequency. If
the oscillation period falls close to the frequency separation used for
the foreground subtraction, incomplete removal of Galactic synchrotron emission
may leave a residual that masks the EoR signal.

One way to avoid the Faraday rotation problem is to measure total
intensity, described by the Stokes I parameter, using a instrument that
rejects the linearly polarized Stokes components Q and U.  However,
all real instruments have various sources of leakage between these
components. To design these instruments and plan future EoR observations
one needs information on the polarized sky brightness in the EoR band
near 150 MHz.

Recent studies at 350 MHz \citep{2006AN....327..487D} report that the
polarized component has much more structure on arc minute angular scales
than the total intensity.  Polarized structure amounting to several
Kelvin has been reported.  Scaling using a synchrotron spectral index
of 2.6 \citep{2008arXiv0802.1525D}, the polarized structure could be
tens of Kelvin in the EoR band.  This means that instrumental leakage
from Q/U to I could severely limit the ability to remove Galactic emission.

This paper places new constraints on polarized sky brightness in the
EoR range of frequencies near 150 MHz. We find the level of polarized
sky brightness is well below that expected by extrapolation from higher
frequencies.  We discuss possible mechanisms which would lead to this
strong depolarization, and suggest future techniques to differentiate
between them.

\section{The GMRT EoR project}

Our group has initiated an effort to search for 21cm structures at the
epoch of reionization using the Giant Metrewave Radio Telescope (GMRT)
in India.

The telescope consists of 30 antennae of 45m diameter. 14 are
designated ``central core'' antennae and lie  within a 1 km area.
The remaining 16 are along 3 arms of lengths up to 10km, designated
East, West, and South.
The dense layout of the core allows high brightness sensitivity, which
is needed for the search for reionization.  For this experiment, the
150 MHz feeds are used.  These consist of orthogonal pairs of folded
dipoles, backed by a ground plane.  These antennas couple to X and Y
linear polarizations, but the two signals pass through a hybrid
coupler before entering the amplifiers.  For each antenna this results
in a pair of right and left circularly polarized signals which are
amplified, upconverted and transmitted optically to the receiver room.
Later processing allows measurement of the full set of Stokes
parameters.

To attempt the EoR experiment, and for other work at GMRT, we built
a new signal processing system for the telescope, the GMRT software
backend (GSB, Roy et al 2009, in preparation).  This consists of 16
commercial Analog to Digital sampling boards installed in an array of
off-the-shelf computers.  The AD boards have 4 input channels, and are
connected to a common clock and trigger signal, allowing synchronous
sampling.  The sampling system is capable of 32 MHz bandwidth but for
this experiment, only 16 MHz was used.  The passband was defined by
an IF filter.  All 60 signals were sampled with 8 bit precision at
33 MSample/s.  Each AD board transfers the digitized data into its
individual host computer.  These streams are Fourier transformed in
blocks of length 4096 samples at 16 bit precision.  The 2048 complex
Fourier coefficients are then rescaled to 4 bits, and sent over a gigabit
network to correlation nodes.  Each block of 128 frequencies is sent to
one of 16 software correlation nodes.  These products, which we call
visibilities, are accumulated for 1/4 second in 16 distinct gates.
Thus the initial visibilities have a frequency and time resolution of
7.8 kHz and 1/4 second (without accounting for the gates), respectively.

The ``gates'' are essential to our polarization calibration technique.
These gates in time are synchronized so each covers one of 16 segments
of the pulsation cycle of a pulsar in the field. The pulsar is a known
source of polarized emission. By comparing pulsar-on to pulsar-off we
can measure the system gain directly using a sky source.  The raw 1/4
second averaged visibilities were stored on disk, as
well as 16 fringe-stopped gated visibilities integrated for 16
seconds, and averaged over frequency into 128 frequency channels.

All these signal processing calculations occur simultaneously, and are
structured as individual asynchronous pipeline processes.  The processor
and network speeds are sufficient that each calculation is completed in
real time and there is less than 10\% data loss.

\section{Data}

Our target field was centered on pulsar PSR B0823+26 (J2000: 08 26
51.38 +26 37 23.79).  The field was chosen to contain a bright pulsar,
and minimal other bright sources.  The galactic latitude is $b=30$ degrees,
and happens to be a relatively cold spot on the sky.  

The brightest radio source within the primary beam, and our primary
flux reference is 5C 7.245, located 12 arc minutes from the pulsar.
Its flux is estimated to be 1.5 Jy at 151 MHz from the 7C Lowdec
Survey \citep{1996MNRAS.282..779W}.  It is a radio galaxy at $z=1.61$
\citep{2001MNRAS.324....1W}.  The source has two components separated
by 16 arc seconds, and is resolved by the longest baselines at GMRT.

A bright source outside the main beam, 3C200, is 2.5 degrees to the
north.  It is attenuated by the primary beam profile, but shows up as
the strongest source in the raw polarized maps due to I to P leakage
in the side lobes.

In Table~\ref{tab:source} we list the 10 brightest sources in the
field after the primary beam power attenuation effect from the 7C
Lowdec Survey.  Also listed are their fluxes from the VLSS survey at
74 MHz \citep{2007AJ....134.1245C} and the Texas survey of radio
sources at 365 MHz \citep{1996AJ....111.1945D}.  Based on these three
surveys we calculate the spectral indices and the errors, as listed in
Table ~\ref{tab:source}.
\begin{table*}
\begin{tabular}{l||cccccccc}
Source & RA [deg] & Dec [deg] & F$_{7C-BEAM}$ [Jy] & F$_{7C}$ [Jy] &
F$_{74}$ [Jy] &
F$_{365}$ [Jy] & $\alpha$  & $\Delta \alpha$ \\
\hline
3C 200 & 126.855 & 29.3128 & 2.135 & 13.435 & 26.78 & 6.407 & 0.93 & 0.07 \\
5C 7.245 & 126.487 & 26.7331 & 1.511 & 1.536 & 2.95 & 0.731 & 0.89 & 0.04 \\
B2 0825+24 & 127.172 & 24.6107 & 0.992 &  2.922 & 10.58 & 0.671 & 1.77
& 0.08 \\
B2 0819+25 & 125.560 & 25.6421 & 0.951 & 1.701 & 3.15 & 0.845 & 0.84 & 0.04  \\
PKS 0832+26 & 128.798 & 26.5764 & 0.706 & 2.124 & 3.43 & 1.384 & 0.62 & 0.10 \\
B2 0829+28 & 128.082 & 27.8792 & 0.682 & 1.634 & 2.64 & 1.030 & 0.63 & 0.08 \\
5C 7.111 & 124.785 & 26.2457 & 0.670 & 1.786 & 3.44 & 1.086 & 0.82 & 0.20 \\
B2 0832+26A & 128.770 & 26.1785 & 0.638 & 1.960 & 3.48 & 1.156 & 0.75 & 0.11 \\
B2 0828+27 & 127.841 & 26.9659 & 0.625 & 0.889 & 1.63 & 0.400 & 0.87 & 0.03 \\
B2 0816+26B & 124.819 & 26.7014 & 0.569 & 1.417 & 2.82 & 0.657 & 0.94 & 0.05 \\
5C 7.223 & 126.028 &  26.4672 & 0.515 & 0.584 & 1.060 &  0.306 & 0.81 & 0.06 \\
\end{tabular}
\caption{The apparent brightest 10 point sources in the observing
 field, listed in descending order of their primary beam attenuated 7C
Catalog flux, listed in the F$_{7C-BEAM}$ column;  F$_{7C}$ indicates
their original flux
 measured from the 7C Radio Survey at 151 MHz, F$_{74}$ from the VLSS survey
 at 74 MHz, and F$_{365}$ from the Texas Survey of Radio Sources at
 365 MHz. $\alpha$ is the spectral index calculated from the three
 surveys and $\Delta \alpha$ the respective errors on the spectral
 indices.
\label{tab:source}}
\end{table*}

This region was mapped at 150 MHz \citep{1970AuJPA..16....1L} and found to have
brightness temperature 170K.
\citet{2008arXiv0802.1525D} have reconstructed a variety of sky maps and 
used them to estimate spectral
indices.  From these reconstructions, we estimate a spectral index of
2.6 for this region, so our sky brightness temperature varies from 200-150K over
our spectral band (140-156 MHz).

Data was taken from Dec 7-Dec 18, 2007, starting at 10PM local time,
going until 7AM.  This period has minimal radio frequency interference.
The data was recorded in one hour ``scans''.  The analysis in the paper
is primarily based on data taken during the night of December 9, which
was fully reduced through the described pipeline at the time of writing
of this paper.

\section{Interference Removal}

Terrestrial Radio Frequency Interference (RFI) was removed in two
stages.  First, spectral-line RFI was flagged.  In each frequency subset of 128 channels,
the distribution of intensities are calculated.  The upper and lower
quartile boundaries are used to estimate the amplitude of the noise.
Any outliers further than $3\sigma$ are masked.
Each mask is 8 kHz wide, and 4 seconds long.  Approximately 1\% of the
data is flagged by this process for deletion.

A bigger problem at GMRT has been broadband RFI.  This is
particularly problematic for short baselines.  
The rationale for the our approach to broadband RFI segregation is that sources on the sky produce
visibilities that oscillate as the Earth turns.  Meanwhile, sources on the ground,
although their intensity typically varies with time, remain at fixed lag for each
baseline. We used a singular value decomposition (SVD) to sort among the terrestrial and celestial classes. 

We assemble the visibilities in a two dimensional rectangular
matrix, $V_i(\nu,t)$ which is a function of baseline $i$, frequency
$\nu$ and time  
$t$.  We find that Earth-fixed broadband sources tend to flicker synchronously in all 
baselines at all frequencies, and are therefore factorizable as $V_{\rm RFI}=L_i(\nu) T(t)$.  We call
$L_i(\nu)$ the {\it visibility template}, and $T_i(t)$ the {\it temporal
  template}.   The RFI
visibility can thus be written as a sum of individual RFI sources $\alpha$,
\begin{equation}
V_{\rm RFI}(\nu,t)=\sum_\alpha L_i^\alpha(\nu) T_i^\alpha(t).
\label{eqn:vrfi}
\end{equation}
Each $L(\nu) T(t)$ product function is a Singular Eigenvector of the matrix.

In contrast, objects on the celestial sphere exhibit fringe rotation at a
unique rate for each baseline, and thus do not factor. Celestial sources
produce very small eigenvalues under a Singular Value Decomposition.

A decomposition of the form (\ref{eqn:vrfi}) can be accomplished via a
Singular Value Decomposition algorithm as long as the eigenvectors are
orthogonal.  A pair of visibility vectors due to two physically separate
sources will tend to have little overlap, since the dimensionality of
this space is huge.  There are of order $30^2=900$ baselines between
antenna pairs, each with four polarization combinations. Two distinct
sources may accidentally fall at the same lag for one pair but will be
separated from each other when other pairings are considered.  The time
templates might not be as orthogonal.  In particular electric arcing from
AC transmission lines, which is likely a the cause of a substantial subset
of the broadband RFI, might occur more often at the AC waveform peak.
This timing can be common to many sources. For this work, only about 100
of the largest SVD eigenvalues are flagged as RFI, in a matrix with about
$10^9$ entries, so even if the eigenvectors are not perfectly orthogonal,
the very sparse space spanned by the first 100 eigenmodes is still given
by an SVD decomposition.

To carry out the procedure we begin by organizing the data in a matrix.
Each hour scan has 14400 time records, which we call the rows of our
matrix, the columns number 2048x60x61 entries which correspond to the
number of frequency channels and baselines. Each matrix entry is a
complex visibility. This 7495680x14400 matrix is then SVD decomposed.
The first 100 right eigenvectors are tagged as ``RFI'', and removed.
This removes 0.7\% of the matrix entries.

We have also used the SVD procedure to physically locate RFI sources.
We selected several of the brightest eigenmodes for further study. Near
field imaging using these eigenvectors has allowed us to localize the
sources on the Earth's surface with 100 meter precision. Visiting these
locations with radio direction finding equipment we have localized
candidate emitters in several cases.  The two brightest appear to
originate from extraneous thin wires hanging vertically from high tension
power lines.  Other sources appear related to intermittent radiation
from transformers. So far, we have visited only a few sources.

The brightest SVD eigenmodes do appear to represent RFI, not celestial
sources, however there are some limitations to the technique.  The
antennas slowly turn as they track objects on the sky and this rotation
leads to phase changes for terrestial RFI sources.  The telescope feed
moves by half a wavelength every 12 minutes.  For an RFI source located
between two antenna, the relative delay changes by a full wavelength.
We believe this sometimes causes individual sources to break up, occupying
several RFI modes. However, as long as the source is bright, it is still
identifiable, even if split into components.

Another limitation: the SVD analysis tends to mis-identify, as RFI,
celestial sources with fringes that rotate slowly.  A baseline that is
oriented perfectly North-South does not produce any fringe rotation,
so baselines with small $v$ are more prone to erroneous projection of
celestial sources onto the RFI modes.  We find that setting the RFI cut
at 100 eigenmodes has a noticeable impact only on baselines with $|v| <
10 \lambda$.  Increasing the number of cut eigenmodes widens the impacted
$v$ range, while substantially decreasing number of cut eigenmodes leaves
significant residual RFI power.

Empirically, a cut at 100 eigenmodes works very well, most visible
broadband RFI had been removed from the data after this process.


\section{Calibration}

Polarization calibration is particularly challenging at low
frequencies.  The primary beam is wide, telescope gain is low,
interference is frequent, the ionosphere is variable and the sky
offers few polarized point sources.  Normally, an iterative procedure
is used, where one starts with a guess for the sky model, determines
the system gain and phases, makes a new map, and uses this as a model.
Errors in the sky model are entangled with errors in the telescope
system calibration.

We used a novel approach by observing a field containing a bright
pulsar.  By subtracting the off-phase from the on-phase, one can lock
in to the modulated emission. Of the 16 gates, we took the gate
containing the pulsar on state, and subtracted the average of the two
flanking gates.  After this subtraction, the measured visibilities
contain only the pulsar. Other celestial or terrestrial signals remain
only if they happen to be modulated in sync with the pulsar.  While
the pulsar flux may vary from pulse to pulse, each antenna sees the
same flux, and the pulsar serves as a common source allowing the
relative gain of the 30 antennae to be measured.  The pulsar also
serves as a guide star.  Pulsar fluxes can be variable in time, but
their positions are very stable.

\begin{figure}
\centerline{\epsfig{file=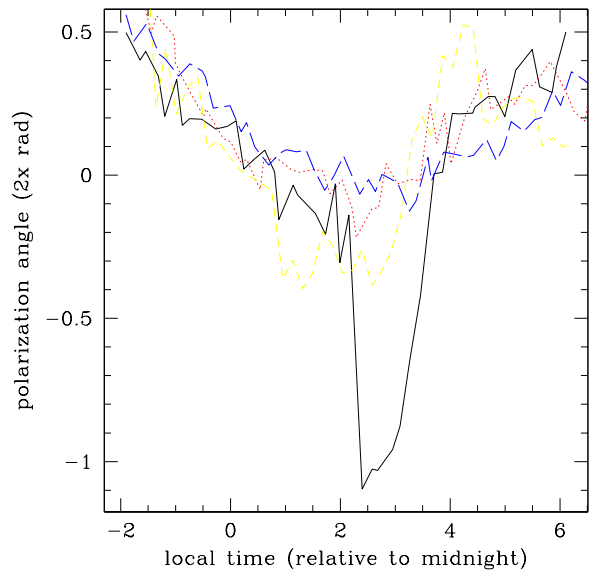, width=\columnwidth}}
\caption{Apparent pulsar polarization angle.  The solid, dotted,
 dashed, long 
 dashed are the mornings of the 18th, 17th, 9th and 16th
 respectively. The vertical axis is twice the polarization position
 angle, i.e. $\tan^{-1}(U/Q)$.  The long
 term change seen in the data (the dip towards the center of the plot)
 is likely due to change in airmass and geomagnetic field angle,
 while the variation between each point and those immediately
 following is consistent with pulsar angle variations at the source.
 The night of the 18th showed a two hour period in which the
 polarization angle departed from the slowly changing behavior
 characteristic of the rest of the data set. This is likely an
 ionospheric anomaly.  For this plot angles are arithmetically
 averaged over one minute intervals, then the median angle for each 9
 minute interval is plotted.}
\label{fig:rm}
\end{figure}

This pulsar is strongly polarized for individual pulses but the
polarization angle varies from pulse to pulse.  The polarization
fraction for individual pulses has been measured to be 40\%
\citep{2008arXiv0802.1202H} at 430 MHz. Many pulsars have polarization
fractions that decrease with frequency so the fraction may
be higher at 150 MHz.  Our data was stored as 16 second averages,
which corresponds to 30 pulses. We found the average polarization fraction for
these averages was 20\%, with an RMS scatter of 8 degrees.  The
polarization fraction decreases as longer averages are assembled.
This is consistent with a highly polarized pulse whose angle varies
from pulse to pulse, with only a slight alignment preference.  This
random angle behavior is also observed at 430 MHz.  Figure
\ref{fig:rm} shows the 9 minute averaged angle, after correcting for
parallactic angle rotation.

We also used the GMRT noise injection system to measure the gain of
the electronic signal chains.  At each prime focus a noise waveform is injected
into the two polarization channels, modulated at 0.5 Hz. We used the 
noise source on a setting which injects 40 K, common between the L
and R channels.  We measured the strength of the noise source in the
L-R correlation of each antenna and used this as a gain reference.

The goal of the calibration process is to determine two complex gains
for each of the 60 signals.  One is the response to its own
polarity, and the other is the leakage from the orthogonally polarized
mode.  The pulsar model appears to work well, with typical RMS phase
closure errors around 1 degree averaged over 20 minutes.  This model
reduces the 1830 complex visibilities into 120 complex gains for each
frequency channel.  The visibilities are reproduced from this reduced
degree of freedom model.  This is called the 'bandpass calibration'.
In addition, the model allows for 1792 complex gains, one for each time
interval of approximately 16 seconds, corresponding to 8 hours of integration.  
The model assumes that the
gains factor as a function of time times a function of frequency.  The
1792x1830 numbers at each frequency are reduced to 1792+120.
Polarization calibration was achieved by using the parallactic angle
rotation to decompose the pulsar into three components: the
``unpolarized'' source which is invariant under parallactic angle
rotation, and two rotating components.  More details can be found in
Pen et al (2008).

The system is fully calibrated using only the pulsar and noise
sources, except for an overall scale described below.  We find that
the unpolarized pulsar flux varies by about 2\% when averaged over one
hour, relative to the average of all noise injection sources.  The noise source
and system gains appear stable for most antennae, with a few
exceptions.  The pulsar flux can be readily subtracted from the time
averaged data.

This calibration procedure results in a polarization map referenced to
the pulsar polarization angle at each frequency.  We then correct this
with the pulsar's known RM=5.9 rad m$^{-2}$ \citep{1974ApJ...188..637M,
2005AJ....129.1993M} for the rest of this paper.

\section{Flux Scale}

We use the pulsar for relative antenna calibration, 
but not to establish the overall gain, since the pulsar flux varies with
time. To accomplish
the overall flux calibration we use four well-measured bright continuum radio sources in the field. 
We estimate the flux of the
four brightest sources that lie within the full-width-half-max (FWHM)
of the primary beam using the estimated spectral indices of these sources from the
catalogs listed in Table~\ref{tab:source} and correcting for the
primary beam power attenuation. We calculate the source flux values
from our data at 151 MHz and match the average flux to that of the 7C
fluxes.  The estimated flux and respective errors are listed in
Table~\ref{tab:flux}. The resulting fractional errors on these four
source fluxes are less than 13\%, and the error on the flux scale in
the calibration procedure is 6\%.  

The calibration of the power spectrum
amplitude depends on the point source flux
calibration described above and also on the primary beam width and profile. 
Throughout, we approximate the primary beam as a round
Gaussian beam with a FWHM of 3.3$^\circ$. This profile is a good fit to our
holographically-measured primary beam power out to about 2 degrees radius.

\begin{table*}
\begin{tabular}{l||cccccc}
Source & RA [deg] & Dec [deg] & F$_{7C}$ [Jy] & F$_{151}$ [Jy] &
Error  \\ 
\hline
5C 7.245 & 126.487 & 26.7331 & 1.536 & 1.74 & 0.13  \\
B2 0819+25 & 125.560 & 25.6421 & 1.701 & 1.51 & 0.11 \\ 
B2 0828+27 & 127.841 & 26.9659 & 0.889 & 0.93 & 0.05  \\ 
5C 7.223 & 126.028 & 26.4672 & 0.584 & 0.53 & 0.09 \\
\end{tabular}
\caption{Sources in the observing field that are used to estimate the
 overall calibration flux scale.  F$_{7C}$ is again the flux from the
 7C catalog at 151 MHz, F$_{151}$ is the flux estimated from our data
 at 151 MHz, and error is the respective fractional error on the
 estimated flux.
\label{tab:flux}}
\end{table*}

\section{Modelling}

After the primary calibration, the dominant polarized source is the
pulsar.  The second brightest apparent polarized source is 3C200, 2.5
degrees to the north, with an intrinsic flux of ca 13 Jy at 150 MHz.
The source is probably not significantly intrinsically polarized, but
rather because the side lobes are strongly polarized, it leaks into the
polarized maps.  

\begin{figure*}
\centerline{\epsfig{file=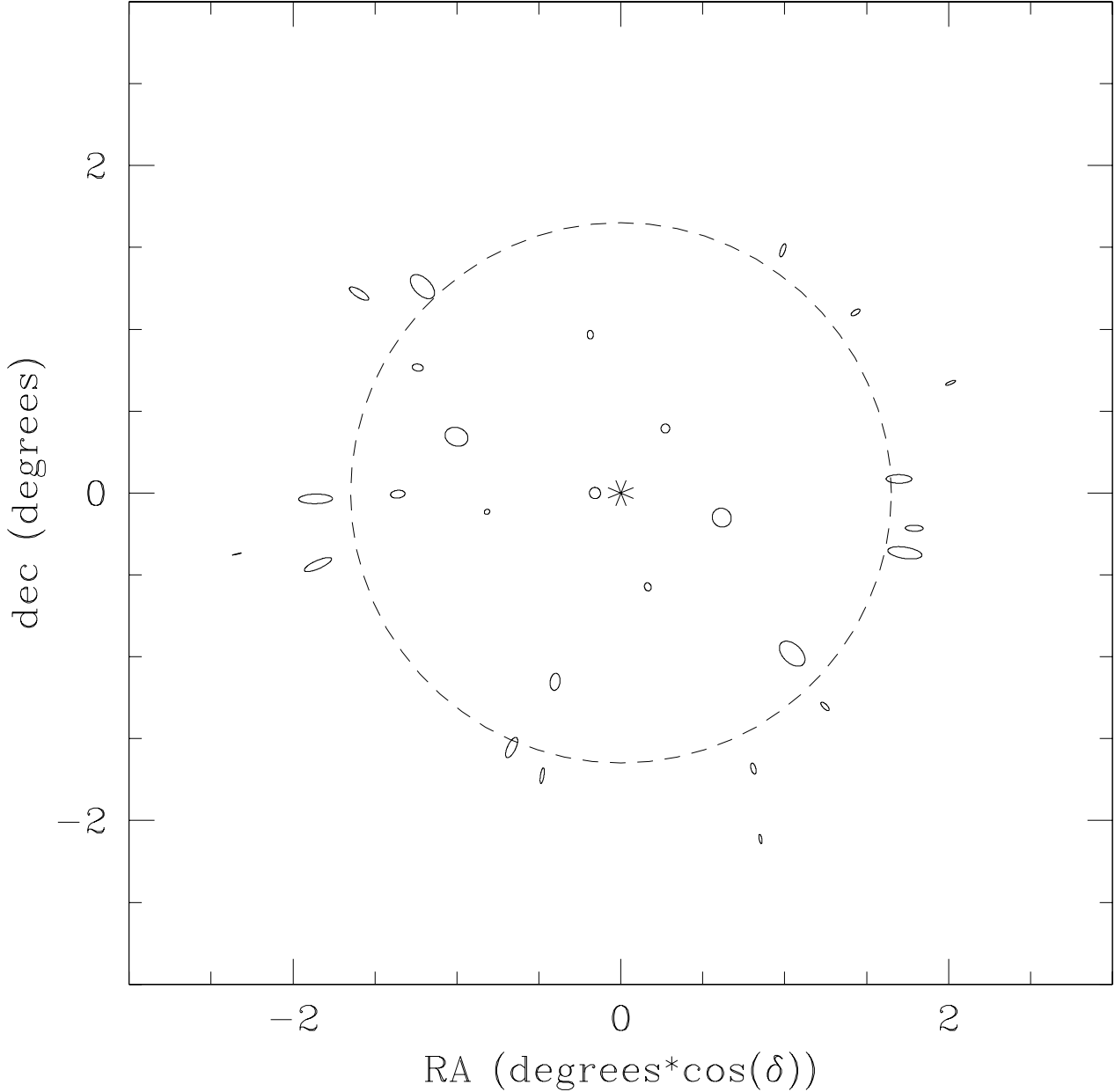, width=\textwidth}}
\caption{
Polarized point source map at Faraday depth $\phi=0$ rad/m$^2$.  The
radius (mean of semi-major and 
semi-minor axes) of each ellipse is proportionate to apparent flux.
Ellipticity is proportionate to polarization.
Map is averaged over 16 MHz.
The 
asterisk in the center is the pulsar.  5C7.245 has been omitted for
clarity. 
The ellipticity is plotted eight times
larger than the apparent polarization.  The global polarization angle
was rotated to align the pattern radially.  The feed support
structures could lead to such a polarization pattern (see text).  The
dashed ring is the half power point of the primary beam.  Sources
outside that radius appear more strongly polarized.
}
\label{fig:stokesI}
\end{figure*}

\begin{figure}
\centerline{\epsfig{file=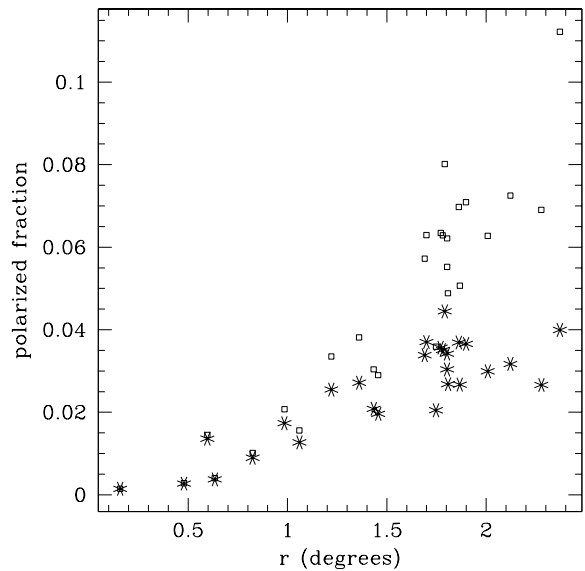, width=\columnwidth}}
\caption{
Apparent polarization of sources at Faraday depth $\phi$=0 rad/m$^2$.  
The small squares are the apparent 
polarizations.  The asterisks are multiplied by the primary beam,
which is an estimate of their impact to the total polarization in the
image.  We see that most of the apparent polarized flux contribution is from
the periphery of the beam.
}
\label{fig:radpol}
\end{figure}

Figure \ref{fig:stokesI} shows the apparent
luminosities and polarizations of the brightest point sources in the
field.  Objects appear more polarized as they are located away from
the image center, suggesting the primary beam as the origin of this
apparent polarization.  Figure \ref{fig:radpol} shows the radial
apparent polarization.

Precise cleaning of the polarized image shown in Figure \ref{fig:stokesI}
will require detailed knowledge of the off-axis leakage at GMRT,
which we have not yet measured.  Despite this imperfect knowledge we 
have been able to make some progress using a modified
version of a standard `clean' procedure. We fringe stop the data
iteratively to sources in descending order of flux as predicted from
the 7C catalog convolved by the Gaussian primary beam model.  At each source
position, we measure the equivalent point sources flux by summing all
baselines.  The point source is modelled as two complex numbers, one
for the intensity and one for polarization. Separate values are fit
every 4 MHz and every 20 minutes.  A point source with this flux is then
subtracted from the data.  We repeat this for the 31 brightest sources
in the field.  For the short baselines, sources are confused,
so we downweighted the central square antennae by a
factor of 10.  A baseline containing one central square antenna gets
10x less weight, and a baseline with two such antennae is downweighted
by 100x.  Similarly, we downweight the last two antennae on each arm,
since some of the sources are resolved by them.  This effectively uses
the intermediate baselines to solve the model, and uses the cross
correlations from arm antenna to central square to extrapolate the
flux on the shortest and longest baselines.

Each of the 31 point sources is fitted with 80 parameters: 4 frequencies by
20 temporal fluxes (one per 20 minutes, total of 400 minutes), for a total
of 248 degrees of freedom.  
To quantify the amount of spurious signal
removal, we added a 30 mJy trace in the form of a Gaussian footprint
centered at u=100, v=-100 with a RMS of $\sigma=20$ in the Fourier space.  Applying the same CLEAN
on this data, and subtracting the original CLEAN data recovers the
original footprint to better than 99\%, while leakage into originally empty
baselines acquires up to 6\% of the peak flux.
We conclude that that this process has at most a small impact on the
flux at $|u|\sim 140$, i.e. $l\sim 900$.

\section{Rotation Measure Synthesis}

At low frequencies, Faraday rotation can have a large effect.  The
pulsar has a rotation measure of RM=5.9
rad/m$^2$.  The apparent
polarization angle rotates as $\chi={\rm RM} \lambda^2+$constant.  We use units
of radians per meter squared throughout this paper.  At 2 m wavelength,
the angle changes 5.3 radians over the 16 MHz observed band.  Since
polarization angles are periodic over 180 degrees, this results in
almost two polarization rotation periods.  After pulsar referenced
polarization calibration, all angles are defined relative to the
pulsar.  A direct 2-D map would strongly suppress sources whose RM is
more than $\Delta$RM$>3.5$ away from the pulsar's, since the rotating
polarization angle causes strong cancellation of the source flux.

To make maps at various Faraday depths, we apply a method known as
rotation measure synthesis \citep{2006AN....327..545B}.  The frequency
channels are mapped into a 3-D grid of $u-v-\lambda^2$.  We chose a
2048x2048x128 grid, where all our frequency channels are scaled to fit
in the first 64 $\lambda^2$ channels.  Padding with 64 zeros increases
the sampling in Faraday depth and results in several reconstructed
slices per RMTF width (although note that this does not affect the
width of the RMTF).  A 3-D FFT then makes a 3D map with axes
$(x,y,\phi)$ The frequencies are binned to the nearest grid point,
resulting in a $2/\pi=64\%$ bandwidth depolarization at the Nyquist
limit of $\phi=56\,$rad$\,$m$^{-2}$.

At Faraday depth $\phi=0$, we can clearly see the beam squint: point sources in the side
lobes appear radially polarized. Our calibration procedure equalizes the polarized response at
the pulsar position, but the leakage will vary across the beam in a
quadrupolar pattern, producing the radial ellipses in Figure \ref{fig:stokesI}.
This squint, a difference between the width of the 
E and H plane antenna patterns, can arise from a variety of reasons. For example,
the prime focus instruments are mounted on four support legs, which are
each narrower than our observing wavelength (2m).  They carry reflecting
currents more efficiently along the legs than transverse.  The illumination
pattern in one linear polarization has a horizontal line blocked, while
the orthogonal polarization pattern would have the opposite line blocked.
To estimate the size of such a squint term we assume of 
order one wavelength is removed, from a 45m dish.  The effective area of the dish 
is about 1000m$^2$, of which 10\% is blocked, and this blockage is different for the two
polarizations.  This could make the E and H beams on the sky different.  The blockage
attenuates the main lobe of the beam, while generating an extended pedestal in the pattern.
In such a model, one expects unpolarized objects within the first null of the primary beam
to appear radially polarized (in terms of the electric field).  Beyond
the null, objects would be tangentially polarized.
Another source of beam squint occurs because the illumination of the dish by the feed
is not exactly round.  

In principle, one could measure the primary beam for both polarizations
and correct the leaked
maps to produce a clean Stokes I map.  This would require a non-local calculation, both on the
sky and in Fourier space, and would require a the use of full matrix inversion.
Codes like CBI-GRIDR \citep{2003ApJ...591..575M} could be adapted for
this purpose.  

%

\section{Results}

We compute the angular power spectrum at different rotation measures. 
The visibilities were directly binned into a power spectrum assuming
a 3.3 degree Gaussian beam.  In this case,
\begin{equation}
\frac{l^2 C_l}{2\pi} = \left(\frac{|u|}{11.9}
\frac{3.3^\circ}{\theta_b}\right)^2 |V(u)|^2 
\label{eqn:cl}
\end{equation}
where $V$ is the visibility in Jansky for a baseline $u$ measured
in wavelengths, and $\theta_b$ is the FWHM of a Gaussian beam.
The left hand side is measured in K$^2$. 
It follows that $l=2\pi u$. 

In order to minimize the impact of non-coplanar w terms, we only used
baselines with $|w|<200$.  We convolved each visibility with a
Gaussian of RMS $\sigma=23$ in the u-v space to reduce the field
of view to the central square degree, which further reduces the impact
of non-coplanarity, and the beam squint.  The phase errors from
non-coplanar $w$ terms are $\Delta \phi \sim w \Delta \theta^2/2$, and
we expect phase errors up to a few degrees at the edge of the reduced
field.

A given point in u-v space is typically covered by several baselines,
at different frequencies.  The RM synthesis maps are a sum of these
points, weighted by the phase factor that accounts for the oscillation
due to the chosen rotation measure.  Power spectra can be used to
summarize the statistics of these maps (Appendix~\ref{app:convention}).
There are several types of power spectra that can be computed, both
2-D and 3-D, containing different types of information.
Perhaps the most straightforward is the square of each gridded
visibility in a  frequency interval $\nu$,
\begin{equation}
P^2=  \frac{l^2 C_l}{2\pi} = \sum_{u,\nu} \left(\frac{|u|}{11.9}
\frac{3.3^o}{\theta_b}\right)^2 |V_{\rm LR}(u,\nu)|^2.
\label{eqn:clu}
\end{equation}
$V_{\rm LR}=Q+iU$ is the complex L-R visibility which contains both Q
and U parameters.  This is a conservative
overestimate of true polarization on the sky, since this statistic
includes thermal noise, polarization leakage, RFI, etc.  By
construction, this power is positive definite. We measured in  frequency
bins $\Delta \nu=1$ MHz, so Faraday rotation within a bin is not expected to
play an important role. 

Since different baselines see the sky at different parallactic angles,
one can also measure the cross correlation between distinct
visibilities.  To reduce the impact of leakage, RFI and thermal noise,
we can discard the self-terms in the power spectrum.  In equation
(\ref{eqn:clu}) we note that the visibility location $u=u(b,t)$ is a
function of the label of baseline $b$ and the time of observation $t$.
Each gridded visibility $|V(u)|^2=\sum_{b,b',t,t'}
V(u(b,t))V^*(u(b',t'))$ can contain contributions from different
baselines taken at different times.  Instead of summing and squaring,
we can square, and drop all contributions where $b=b'$.

\begin{figure*}
\centerline{\epsfig{file=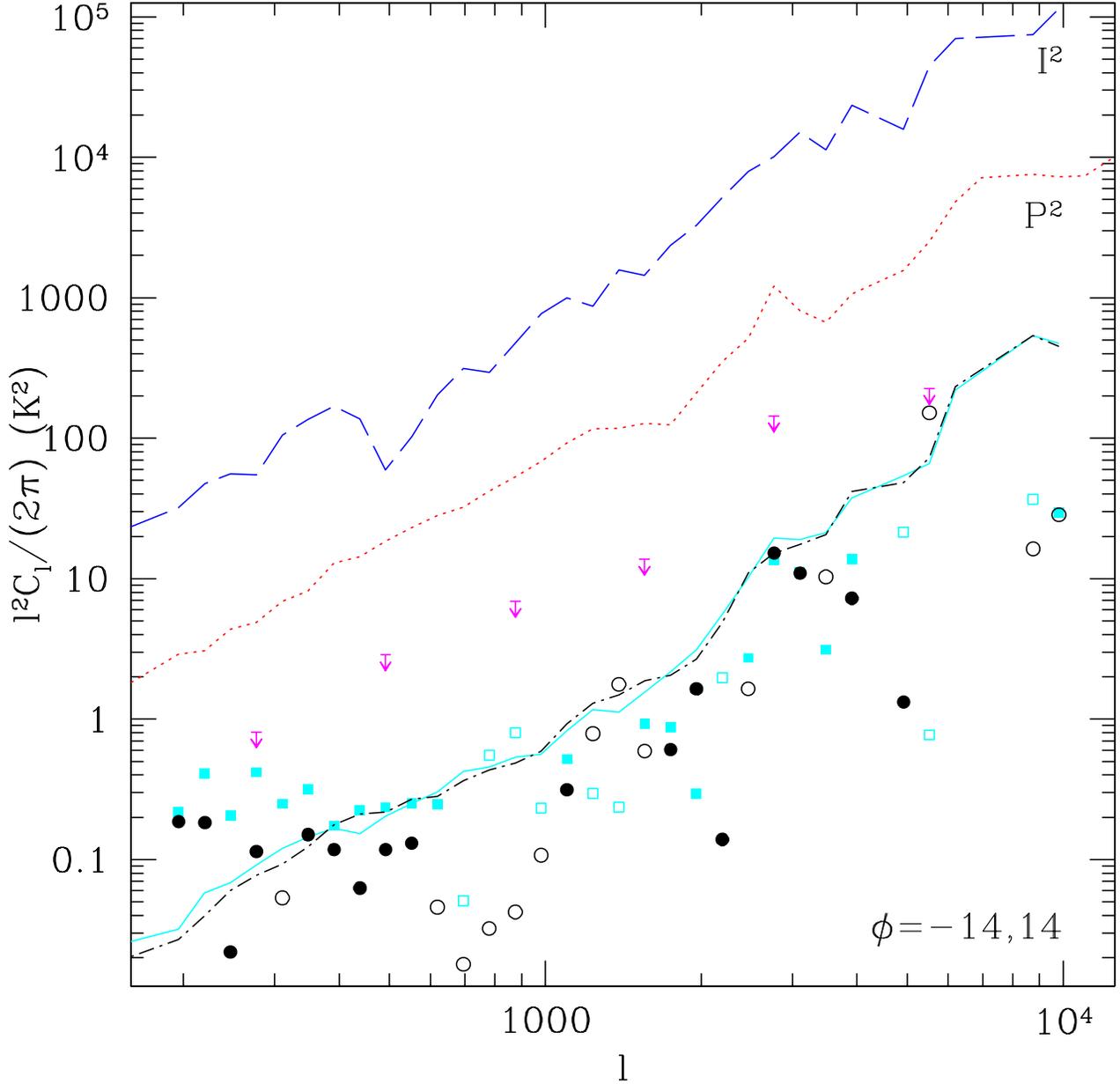, width=\textwidth}}
\caption{Power spectrum of unpolarized ($I^2$, long dash) and
total polarized emission  ($P^2$, dotted) over the central square
degree, after subtracting the 31 brightest sources in the primary
beam.   This polarized signal  
includes leakage and thermal noise.  The upper bound symbols are a
2-$\sigma$ estimate of polarization bound using baseline cross
correlation.  They are averages of the polarization power in 1 MHz bands.
The lower solid and dot-dashed lines are the total
power polarization measured at
$\phi$=-14 and 14 over a width $\Delta\phi$=3.5, as
determined from the map power.  
Using the cross correlation technique described in the text, we plot
a less conservative estimate with circles for $\phi=-14$ and squares
for $\phi=+14$.  Open symbols denote negative power, and filled are
positive values.  For details on the unit conversions, see the appendix.
}
\label{fig:cl}
\end{figure*}

Computationally it is expensive to store all baselines contributing to
a u-v grid point, so we randomly separate baselines into two bins,
grid the visibilities from each bin, and compute their cross
correlation.  We repeat this many times to include all possible pairs
of baselines contributing to a product, in this particular case 129.
This results in an average power $\langle C_l\rangle$, as well as a
standard deviation $\delta C_l$.  The latter is a rough estimate of
the standard deviation on $\langle C_l \rangle$.  The variance on the
mean is smaller than the variance on each subsample.  If the baselines
were densely distributed, the variance is overestimated by a factor of
$2$.  The actual baselines are not dense, so sometimes one of
the elements of the pair will be empty. This causes the variance to be
further overestimated.
The average $\langle C_l \rangle$
appears consistent with zero, so we roughly interpret $\delta C_l$ as an
effective 2$\sigma$ 
upper
bound, shown in Figure \ref{fig:cl}.

If one fixes a rotation measure, all frequencies can be summed into a
single u-v grid point, which further improves sensitivities, and localizes
the RM of the power spectrum.  
Before subtracting the point sources, excess power is
visible at $\phi=0$ when imaging a 4 degree field. After
point source subtraction, we saw no rotation measure at which there is a
statistically significant signal.  Restricting the field to one degree
using a primary beam convolution in u-v space should further suppress
any potential contamination.  We apply $\theta_b=1$ degree in Equation
(\ref{eqn:clu}).  We show two representative lines in Figure \ref{fig:cl}
at $\phi=-14$ and 14 rad m$^{-2}$, where we have plotted the absolute value of the power.
Because we discarded auto-correlation terms, noise can lead to positive
or negative correlations.  The effective width of the window
in Faraday measure space is determined by the 16 MHz bandwidth, which corresponds to
bins of width $\Delta\phi\sim3.5\,$rad$\,$m$^{-2}$.

[Note: there are several ways to measure the width of the RMTF. For a 
top-hat range of frequencies, $|$RMTF$|(\phi)=|\sinc(\phi\Delta\lambda^2)|$
(the phase depends on the wavelength to which one de-rotates). The 
width relevant for power spectrum analysis (Appendix~\ref{app:convention}) 
is $\int|$RMTF$|^2\,{\rm d}\phi = \pi/\Delta(\lambda^2)$ since if a slice is 
suppressed by the RMTF then its power spectrum is suppressed by RMTF$^2$. 
One could alternatively compute the FWHM of $|$RMTF$|$, which for the sinc-function is 
3.79/$\Delta(\lambda^2)$.]

The observed apparent polarized power could either be polarized
structures on the sky, or residual leakage from beam squint.  Towards
the edge of the beam, point sources appear in the polarized maps at up
to 10\% of their unpolarized flux, with a quadrupolar pattern
characteristic of beam squint: objects on two diametric sides of the
fields appear as positive sources, and at 90 degrees appear negative.

\begin{figure*}
\centerline{\epsfig{file=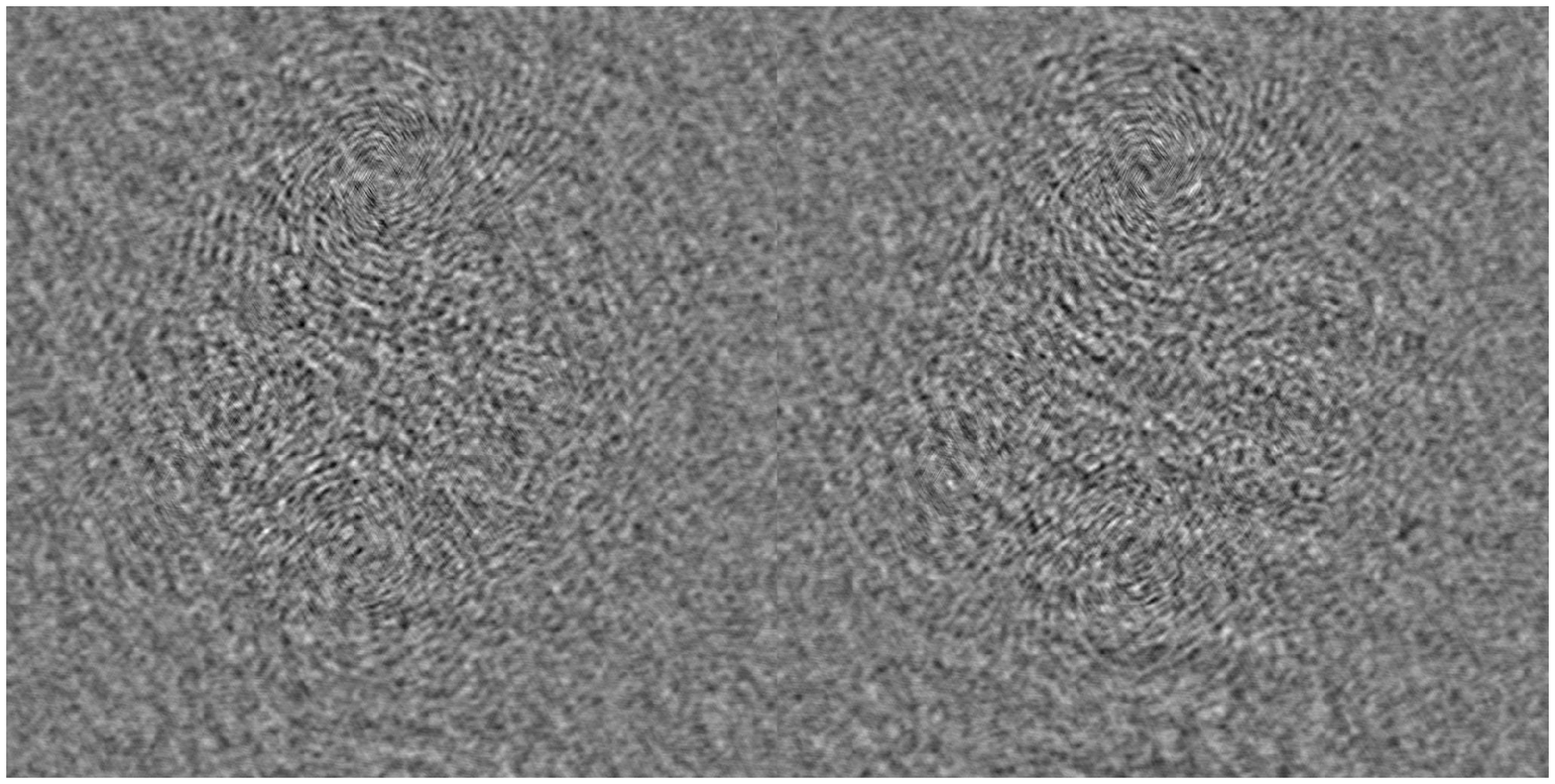,width=6in}}
\vspace{-3.21in}\hspace{-0.24in}
\epsfig{file=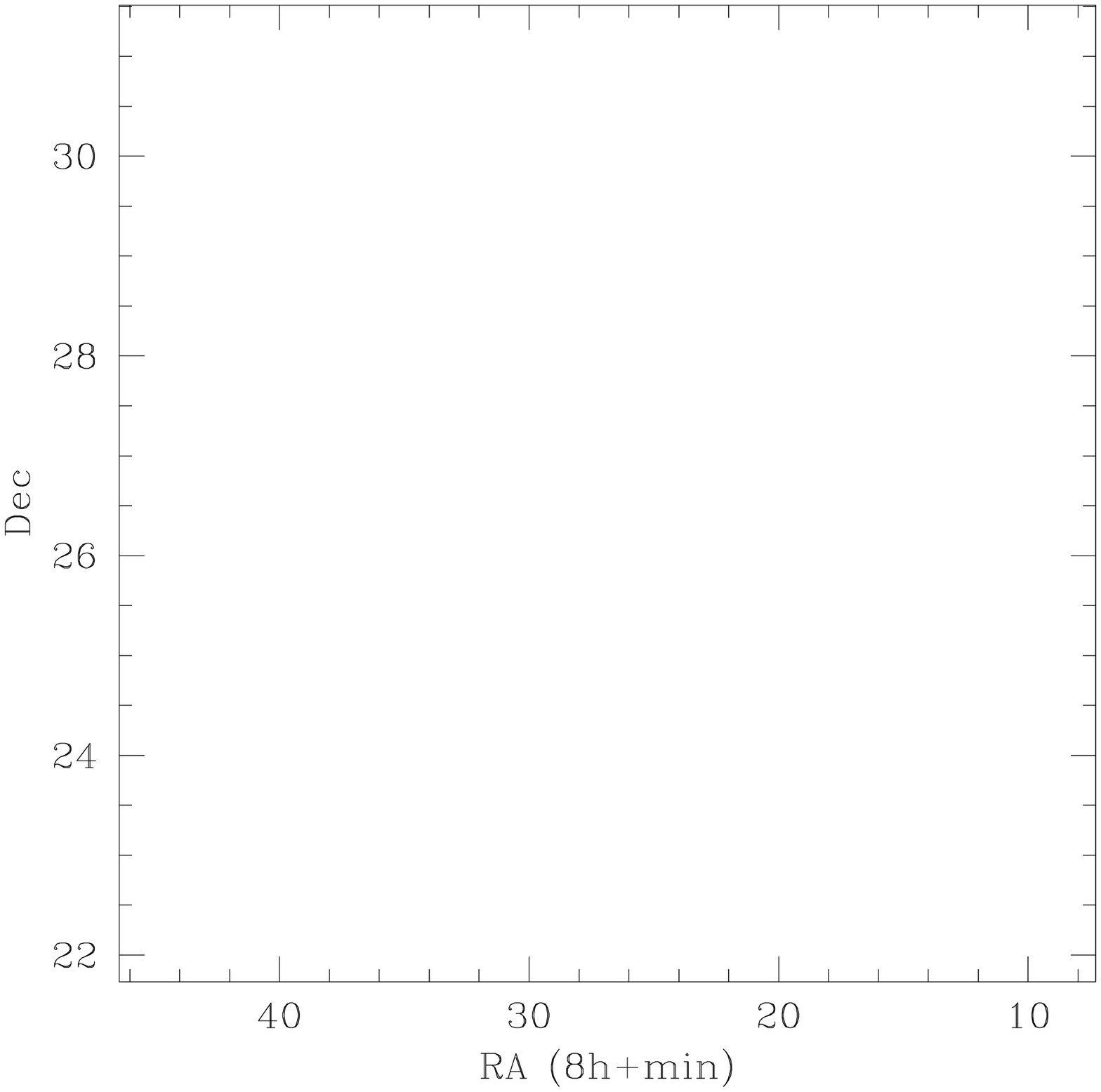,width=3.58in}\hspace{-0.58in}\epsfig{file=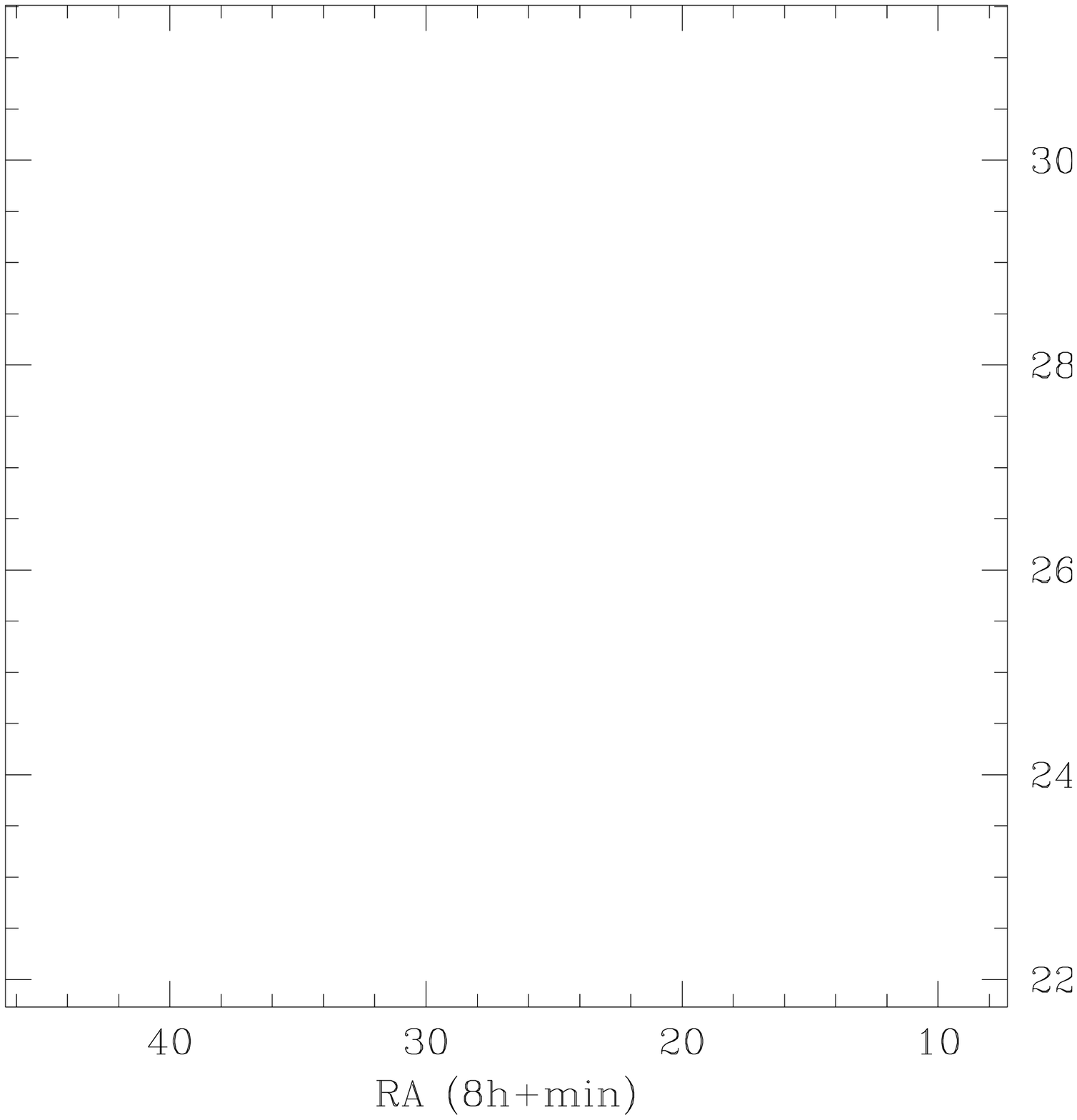,width=3.58in}
\caption{Polarized sky image at $\phi=3\,$rad$\,$m$^{-2}$.  The left panel is the Stokes Q
  intensity, while the right is Stokes U.  All baselines with $|w|<200$
  are included.  The field of view is 9.78 degrees, which is three
  primary beams.  Real polarized sky structure as well as leakage I into  Q,U 
  will have higher contrast within the main beam region. Sources that generally have the same contrast
  inside and outside the main beam area include RFI and the aggregate extended image artifacts of multiple
  bright sources far outside the main beam. These images do show a weak enhancement of contrast in the main 
  beam region, but the enhancement is too weak to allow confidence in detection of sky structure. 
  For this reason we interpret the measured power spectra from these images as upper limits to
  the polarized sky structure. 
  Artifacts from one known bright source 3C200 are faintly visible as concentric rings centered on 
 the source location. 3C200 is in the upper center of the image just outside the main beam.
  }
\label{fig:stokesqu}
\end{figure*}

We show a map of the point-source-subtracted sky in Stokes Q and U in
Figure \ref{fig:stokesqu} at $\phi=3\,$rad$\,$m$^{-2}$, half way to the pulsar RM.  Excess
power is still discernible in the region of the primary beam.  The
point source subtraction reduces the impact of this leakage, but it
could still leave residuals.  Our apparent measured polarized signal might not
be truly polarized on the sky, but rather leakage from the unpolarized
sky brightness or residual man-made interference. 
We treat the measured power in this image as an upper limit
to the polarized sky brightness.

\subsection{3D Power Spectrum}

EoR structure is expected to be three-dimensional (3D), so 
we need to compare the 3D power spectrum of the polarized foreground to
that of the expected signal.
Theorists have traditionally written the
EoR power spectrum in terms of the temperature variance per logarithmic
range in $k$ \citep{2004MNRAS.352..142B, 2005ApJ...624L..65B}:
\begin{equation}
\Delta^2_T(k,\mu) = \frac{k^3}{2\pi^2}P_T(k,\mu),
\end{equation}
where $\mu$ is the cosine of the angle between the wave-vector and the
line of sight and $P_T$ is the 3D temperature power spectrum.  It is also common to work in terms of the radial and transverse 
components of the wavenumber, $k_\parallel=k\mu$ and $k_\perp = k\sqrt{1-\mu^2}$.

It is appropriate to evaluate EoR foregrounds by projecting the Galactic polarization structure
onto the cosmological volume using the usual mapping $\nu \rightarrow z=1420\,$MHz$/\nu-1$, and
estimating the polarized 3D power spectrum that one would infer.
Details of this power spectrum definition are given in Appendix~\ref{app:convention}.
At our central frequency $\nu = 148\,$MHz, and with the WMAP determination of $\Omega_m=0.26$ \citep{2008arXiv0803.0586D}
the constants described in the appendix are: the redshift $z=8.59$; comoving radial distance $D=6720h^{-1}\,$Mpc
(where $H_0=100h\,$km$\,$s$^{-1}\,$Mpc$^{-1}$); $H/c=5.04\times 10^{-3}h\,$Mpc$^{-1}$; and $C=12.9h^{-1}\,$Mpc$\,$MHz$^{-1}$.
The mapping of Eq.~(\ref{eq:phi-k}) between the Faraday depth and the apparent radial wavenumber is
$k_\parallel = 0.0086h\,$Mpc$^{-1}\,($rad$\,$m$^{-2})^{-1}\,\phi$.
The 3D polarized power spectrum $P_P(k,\mu)$ is simply related 
to the 2D power spectrum $C^P_l$ of the Faraday depth slice in the RM synthesis map at this value of $\phi$.
One can then obtain
\begin{equation}
\Delta_P^2({\bmath k}) = \frac{k^3}{2\pi^2}P_P(k).
\end{equation}

\begin{figure}
\centerline{\epsfig{file=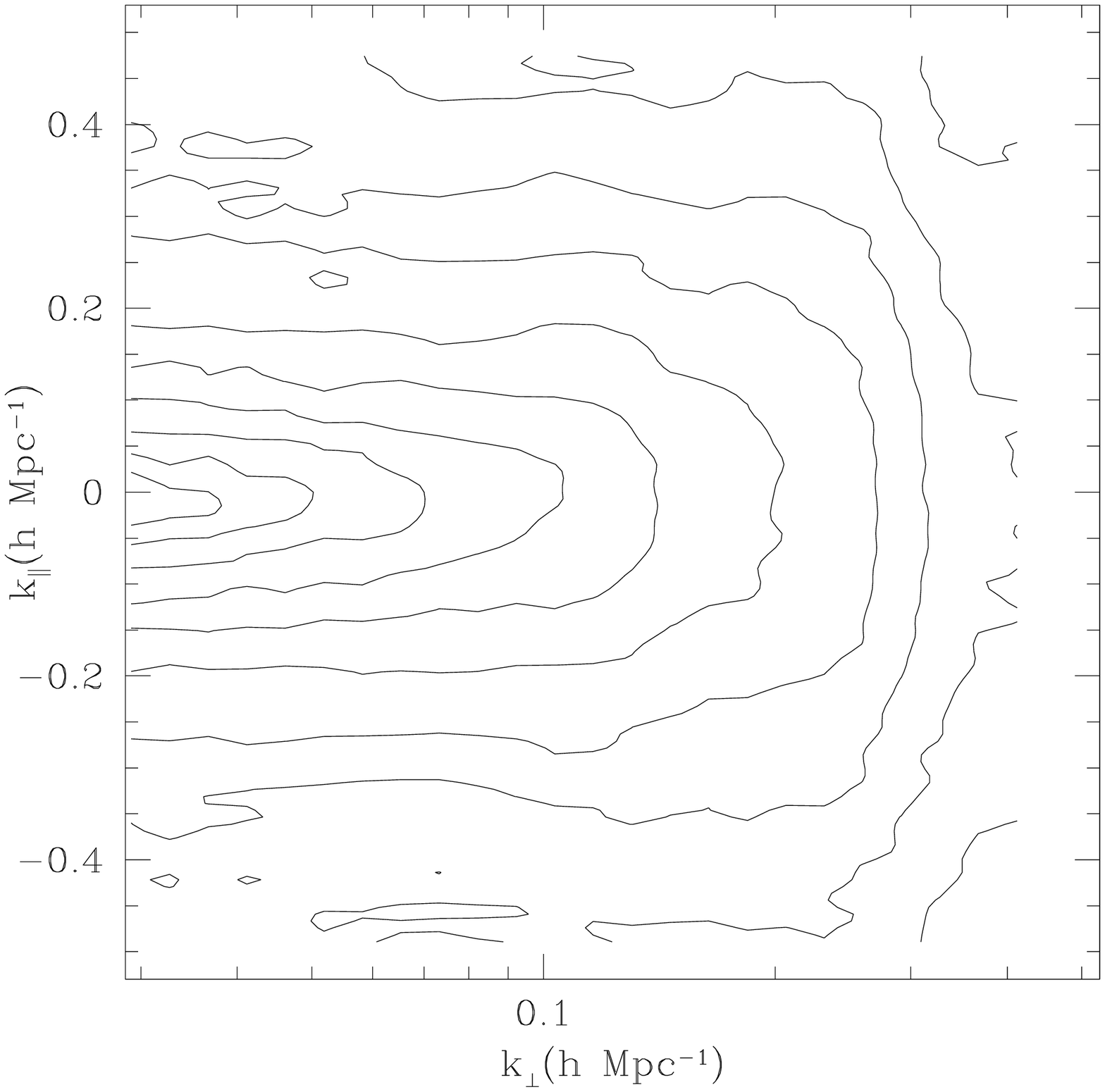,width=3.5in}}
\caption{The 3D polarization power spectra based on our observed
visibilities.  These are estimated from the power of maps at fixed
rotation measures.
The transverse scale has been converted from angular to
comoving units at $z=8.5$.  The parallel scale is converted from 
$\phi$ according to Eq.~(\ref{eq:phi-k}).  The contours are of 
$\Delta^2_P$ (see text), and
start at 0.0069 K$^2$ in the left middle near $k\sim 0$, increasing
logarithmically by 10 per contour level.
On large scales, $k_\perp<0.3h\,$Mpc$^{-1}$
our
observations set an upper limit equivalent to a few Kelvins per $\ln k$ in
3D space.}
\label{fig:3dpow}
\end{figure}

The 3D power spectra $\Delta_P^2({\bmath k})$ are shown in Figure 6;
the power per $\ln k$ is at most $\sim 2\,$K at $k\le 0.1h\,$Mpc$^{-1}$.
This rotation measure corresponds to polarized emission at a line-of-sight
wavenumber of $k_\parallel = 0.12h\,$Mpc$^{-1}$.  The implied 3D power
spectra are shown in Figure~\ref{fig:3dpow}.

\section{Discussion}

If extended polarized emission is bright at 150 MHz, leakage from
Stokes $Q$ or $U$ into Stokes $I$ may make removal of this foreground from
21 cm EoR images quite difficult.  Removal of the synchrotron intensity
foreground from EoR images requires projecting out structure that
varies slowly with frequency (i.e. has $k_\parallel\approx 0$).
This also removes any component of the EoR signal that varies slowly
along the line of sight, i.e. it suppresses modes with
$|k_\parallel|$ less than a few times $(C\Delta\nu)^{-1}$.  The anticipated
EoR structure is at scales of order $k\sim0.1h\,$Mpc$^{-1}$; this
range of scales that will be contaminated by
$Q,U\rightarrow I$ leakage of
polarized structure at Faraday depth of 
order $|\phi|\sim 10\,$rad$\,$m$^{-2}$
[see Eq.~(\ref{eq:phi-k})].
This is the Faraday depth range for which polarization leakage causes
the most difficulty for an EoR search, although if the polarization
leakage varies rapidly with frequency other values of $\phi$ are
also of concern (see Appendix~\ref{ssa:3dp}).

Previous work at 350 MHz \citep{2006AN....327..487D} indicated
polarized emission with spatial variations of several Kelvin on angular scales
of tens of arcminutes.  This polarized structure is reported to be
brighter than the unpolarized component at the same angular scale.
Extrapolating this polarized structure to 150 MHz with a spectral
index of 2.6, this corresponds to several tens of Kelvin in the EoR band. To remove
this foreground would require that instrumental Q or U to I leakage in
EoR observations be controlled to $\ll 10^{-3}$.

Our observed upper limit to polarized sky structure in frequency channels
of width 1 MHz is in the range of 1--3 K at $300<l<1000$, with weaker upper
limits at smaller scales.
This is shown in
Figure \ref{fig:cl}. For the field we studied the polarized sky structure is an
order of magnitude smaller than one would have predicted by scaling
from the published 350 MHz observations.

The results reported in \cite{2006AN....327..487D} are at different
locations on the sky, so a direct comparison is not straightforward.
The \citep{2006AN....327..487D} field is centered at 13h14m +45
(J2000), while our field is at 08h26m +26.  However, one way to
attempt a comparison is to look at the WMAP measurement of
polarization emission on the two fields \citep{2008arXiv0803.0732H}.

Smoothing the WMAP data to 3.7 degree resolution, 
the 13h field has a polarized temperature of $13
\mu$K, while  our 08h field is $17 \mu$K.  The noise is comparable to
the difference, so we estimate our field has a similar net polarization at
high frequency.
In the Haslam 408 MHz intensity maps \citep{1982A&AS...47....1H}, the 13h field has a temperature of
20K which is similar
to the 08h field at 18K. Using the available data the fields seem quite similar.

\cite{2006AN....327..487D} explain the observed polarization
structures as due to a magnetic field that is smooth across their
field, combined with an electron density that varies. This causes
structure in the Faraday depth allowing the polarized sky structure
to exceed the total intensity variations.  We see no reason to think
our field would be different in this regard.

\citet{1996AstL...22..582V} provide a model of diffuse Galactic polarization toward the North Celestial Pole (NCP).  They find polarization 
$\sqrt{Q^2+U^2}$ of $3.5\pm1.0$\ K at 88 MHz (with a 24$^\circ$ FWHM beam) and $2.15\pm0.25$\ K at 200 MHz (8$^\circ$ FWHM).\footnote{Note that the 
numbers in Table 1 of \citet{1996AstL...22..582V} are actually $2\sqrt{Q^2+U^2}$ due to their normalization convention for polarized brightness.} Their 
result is difficult to compare to ours, for several reasons.  First, there is no overlap in $(u,v)$ coverage between their single-dish (or phased 
antenna array) measurements and our interferometer results (which have no zero-length baselines).  Second, one must interpolate between frequencies to 
compare to our 150 MHz results; \citet{1996AstL...22..582V} can fit physical models with polarization nulls between 88 and 200 MHz, but the number and 
location of such nulls is not uniquely determined.  Finally the observations are in different parts of the sky, and one would not expect 
geometry-dependent properties such as the frequencies of polarization nulls to be exactly the same.

\subsection{Depolarization Models}

Several depolarizing effects have been predicted at low
frequency\citep{1966MNRAS.133...67B,1998MNRAS.299..189S,2004mim..proc...93V}.
These include beam depolarization due to field gradients, or sub-beam
structure, and depth depolarization due to co-location of the rotating
and emitting regions.  We now consider each effect in turn as possible
explanations for the low level of polarization at 150 MHz.

First we consider beam depolarization \citep{1966MNRAS.133...67B}.
\cite{2006AN....327..487D} 
reported RM gradients of $\nabla$RM=2--4 rad$\,$m$^{-2}\,$deg$^{-1}$.
At 150 MHz ($\lambda=2\,$m) a gradient this size would result in 
a periodic modulation of the polarization
vector field at
\begin{equation}
l = \frac{360\,{\rm deg}}{\pi}\,\nabla{\rm RM}\,\lambda^2 \sim 1000-2000.
\end{equation}
We could detect such a periodic variation, but do not see such a
feature.  If the spatial gradients were substantially larger in our
field, the periodic structures could occur on scales smaller than we constrain.
However, this would require a RM
gradient larger than 20 rad$\,$m$^{-2}\,$deg$^{-1}$. Beam
depolarization due to presence of a smooth RM gradient 
seems an unlikely explanation for the low polarization.

Alternatively, beam depolarization could arise if there was a small
scale random magnetic field.  An equipartition magnetic field in the
galaxy could be several $\mu$G, but this is a factor of several larger
than the coherent field.  The coherent field is observed to contribute
an RM=5.9 rad$\,$m$^{-2}$ to the pulsar, which results in a 23 radian
rotation of the polarization vector.  The dispersion measure
DM=19.4751
cm$^{-2}\,$pc \citep{2005AJ....129.1993M,1974ApJ...188..637M}, results
in a mean electron-weighted magnetic field along the line of sight
field of $\langle B_\parallel \rangle=1.232 {\rm RM/DM} =
0.37\mu$G. This coherent part of the field is lower than the
equipartition value making plausible the idea that there is fine
structure in the field.

Substantial beam depolarization will occur if the small scale field
structure produces variation of rotation of order unity across the
synthesized beam.  Given the current sensitivity at GMRT, structures at
$l\gg 10^4$ are unconstrained.  A statistically isotropic field on
scales smaller than that would have a correlation length $10^4$
shorter than the line of sight.  Random rotation measures add
stochastically, so the net angle would be equivalent to a 100 times
weaker, but coherent field.  So a random magnetic field which has five
times the field strength of the coherent field and a correlation
length $10^4$ times shorter than the rotation screen would lead to
beam depolarization.  Taking an electron scale height of 1 kpc
\citep{2002astro.ph..7156C}, which at $b=30^\circ $ is 2kpc in
projection, would correspond to a fine-structure scale of $\sim 0.2
$pc.  A magnetic field of 3 $\mu$G on these scales would suffice to
depolarize, which is in the plausible range of galactic magnetic
fields.  The spatial scales are a bit smaller than the outer scales of
1 pc proposed in turbulence models \citep{2008arXiv0802.2740H}, 
but may still be plausible given the general nature of these arguments.
At our galactic latitudes, the outer scale may also be substantially
larger.

We now take the analysis of fine scale beam depolarization one step
further by assuming a turbulence power spectrum.  We consider a Faraday
thin emitting region with a rotating medium in the foreground.
We model the RM with a Kolmogorov-like power spectrum
on small angular scales, i.e. $\Delta{\rm
  RM}^2(\theta)\propto\theta^{5/3}$ \citep{2008arXiv0802.2740H}.  In
such a model the polarization direction is coherent over patches of
size $\theta_c$ where $\Delta{\rm RM}(\theta_c)\lambda^2\sim 1$.  For
the Kolmogorov spectrum this implies $\theta_c\propto\lambda^{-12/5}$.
The peak of the observed polarized power spectrum then scales as
$l_{\rm peak}\sim\theta_c^{-1}\propto\lambda^{12/5}$.  Observations at
350 MHz \citep{2006AN....327..487D} typically find polarized emission
at scales of order 10 arcmin.  There do not appear to be published
power spectra of these polarization maps but if the $Q$ and $U$ Stokes
parameters change sign on this scale, as one would expect from the
abundant depolarization canals \citep{2004A&A...427..549H}, then these
structures contribute mainly at $l\sim 10^3$.  Scaling to 150 MHz
would move this power up to smaller scales by a factor of
$(350/150)^{12/5}\sim 8$, or $l_{\rm peak}\sim 10^4$.  This reduces
the tension with our observations, which have upper limits on
$(l^2C_l/2\pi)^{1/2}$ of a few tens of K on these scales (see
Fig.~\ref{fig:cl}).  This is close to the expected signal from
scaling from higher frequencies; the additional presence of depth
depolarization, or different turbulence parameters in our field would
eliminate any tension.

Next we consider depth depolarization, which was already discussed
by \cite{1966MNRAS.133...67B}.  This occurs because the emission and rotating
media are mixed, and some rotation occurs within the source.
This effect has been reported at 1.4 GHz for observations looking right into
the Galactic Plane, where rotation measures can exceed 1000 rad$\,$m$^{-2}$
\citep{2003ApJ...585..785U}.  Our regime has a similar product of RM$\lambda^2$,
so it is tempting to draw analogies.
For this part of the discussion we assume 
a uniform magnetic field.  The synchrotron emission arises
from relativistic electrons, which make up a small fraction of the total
electron count.  However, the Faraday rotation is caused by all free electrons.

The strength of the depolarization depends on the profile of the emitting region.
An example of a rapidly depolarizing model is a Gaussian
emission region embedded in a medium with uniform $n_e$ and ${\bmath B}$.   This
might be a reasonable model of an emission region due to cosmic rays
diffusing across magnetic field lines.  Some prominent structures in
the sky, such as the Cygnus loop, may be examples of such sources.
This results in a Faraday dispersion function that is a Gaussian
\begin{equation}
F(\phi) = \frac{P_0}{\sqrt{2\pi}\,\sigma}\exp\frac{-(\phi-\phi_0)^2}{2\sigma^2}.
\end{equation}
The observed polarization vector is
\begin{equation}
P = {\rm e}^{2{\rm i} \phi_0 \lambda^2} {\rm e}^{-2\sigma^2\lambda^4}.
\label{eqn:gaussdepth}
\end{equation}
This model has an exponential suppression with argument $\propto \lambda^4$, so a
factor of $\sim 2$ suppression at 350 MHz would be a factor of $10^9$
suppression at 150 MHz.  Using the dispersion and rotation measures to
the pulsar, an emission structure at a distance similar to that of the
pulsar, with a Gaussian width 10\% of the extent of the rotating
medium would produce an order of magnitude attenuation of the net polarization between 350 and
150 MHz.

However it is unlikely the electron density follows the precise Gaussian 
profile needed to achieve 
the deep cancellation of the polarization from different regions.
We now consider an alternative scenario, which would lead to much less
depth depolarization.  Assume the emission spatial structure traces the
ionized gas structure perfectly. Two scenarios
could lead to depolarization: (i) the ionized and relativistic populations
are uniformly distributed, but the magnetic field is confined to flux
tubes; or (ii) the ionized and relativistic components trace each other
(for example, if they were carried by the same magnetic field).  In both
cases, the Faraday dispersion function is a top-hat, i.e.
$F(\phi) = P_0/|\phi_0|$ between 0 and $\phi_0$.  This leads to a net
polarization
\begin{equation}
P = P_0 {\rm e}^{{\rm i}\phi_0\lambda^2} \frac{\sin(\phi_0\lambda^2)}{\phi_0\lambda^2}.
\label{eqn:mixdepth}
\end{equation}
This leads to an oscillating depolarization with a $\sim \lambda^{-2}$ envelope,
which at 150 MHz gives
an additional factor of only 5 suppression compared to
the 350 MHz data, less than the increase due to the spectral index.  In this model one
would expect 5--10 K polarized emission, which is not observed at $l<10^3$.

The emission region could also be only partially overlapping with the
rotation region. Many complex permutations are plausible.  The simple
models presented above are meant to illustrate that strong depth
depolarization is possible, given the known parameters.

\subsection{Prospects}

To further study the nature of the depolarization, multi-wavelength
polarized observations of the same field are needed.  Repeated
observations during daytime would give ionospheric rotation
measure variations that could separate beam squint from intrinsic polarization.
Much more sensitive polarized maps then could be made.  Such
observation would likely concentrate on fields that show substantial
polarized flux.

Several steps could also be implemented in the future.  Just as
parallactic angle rotation allows us to decompose the polarization
components of the pulsar, one could use time variable ionospheric
rotation measure to unleak the sources.  The ionospheric Faraday rotation
changes through the day-night cycle, as we see in Figure \ref{fig:rm}.
After a few months, the same point on the sky will be viewed through
a substantially thicker ionosphere, and the Faraday depth of all sources
will be modulated by an ionospheric contribution $\Delta\phi$.  The apparent angle
of polarization will change, while unpolarized light stays unchanged.
This allows a separation of polarized and unpolarized components which
does not require modelling of beam squint.  Specifically, in circularly
polarized baselines, a two epoch observations at 1,2 measures
\begin{eqnarray}
V^{\rm LR}_1&=&V^{\rm LR}_{\rm sky} + V^{\rm LR}_{\rm leak} \nonumber\\
V^{\rm LR}_2&=&V^{\rm LR}_{\rm sky}\exp(2i \Delta\phi\,\lambda^2) + 
V^{\rm LR}_{\rm leak}.
\end{eqnarray}
We observe the two visibilities $V_{1,2}$ at the same azimuth and elevation, so
knowing $\Delta\phi$ allows us to solve for the sky and leakage visibilities.

To test the scenario of small angle beam depolarization, one could
compare the rotation measure of pulsars in globular clusters.  A
search of the ATNF pulsar database \citep{2005AJ....129.1993M} did not
turn up any close pairs.  M15 and 47 Tuc have multiple pulsars
separated by arc minutes which would be predicted to have fractional
variations in RM of order 10\% in a beam depolarization scenario.

As far as the search for high redshift reionization is concerned, one
wants fields with low Galactic polarized flux. The upper limit to
polarization we report indicates that suitable fields are available.

Other steps to reduce the polarization leakage could include
mosaicking a wide field, where the angular patterns would cancel, as
was done in the work of \cite{2006AN....327..487D}. In addition, a
raster of observation of the pulsar would allow measurement of the
polarization calibration into the skirts of the main beam.

Given the low level of polarized sky structure we report, along with
the estimates of polarization leakage at GMRT presented in Figure
\ref{fig:radpol}, we can estimate that even a modest polarization
calibration will allow measurement of EoR sky structure.  The one
sigma upper limit to sky structure measured as a 3-D power spectrum
$[k^3P_P(k)/2\pi^2]^{1/2}$ at scales relevant to EoR ($k_\perp\sim
k_\parallel\sim 0.1h\,$Mpc$^{-1}$) is around 2 K, and the I to Q, U
leakage at the edge of the main beam is 0.03.  Assuming that the Q,
U to I leakage is also around 0.03 the polarized sky structure on
this field could create apparent sky structure that would interfere
with detection of the EoR signal at the 60 mK in 3-D. This is the
level of polarization artifact that may occur at the edge of
the main beam.  Leakage is smaller near the center.  A factor of 10
primary beam leakage correction would be sufficient to bring potential
polarized leakage levels below the $\la$ 10 mK expected EoR signal.
Polarization calibration made by rastering the pulsar across the main
beam, along with mosaic observations may allow the beam squint terms to
be measured and corrected for to achieve this additional factor of ten
reduction in the polarized emission artifacts, allowing a measurement
of large scale structure during the Epoch of Reionization.  The required
calibration should be attainable using the pulsar technique.  There are
caveats however; for example, if there is much stronger polarization in
neighboring fields then polarization leakage through the sidelobes or
diffraction spikes may be an issue.  As we have seen in this paper,
bright unpolarized sources outside the beam can appear polarized, but
at zero Faraday depth.

\section{Conclusions}

The GMRT EoR data has resulted in the strongest upper limits to
polarized emission at 150 MHz to date.  The observed polarization
is an order of magnitude smaller than had been expected from higher
frequency observations.  We note that comparison was not made on the
same fields. We estimated the expected beam and depth polarization
effects, but these are model dependent and poorly constrained by the data.
Within the uncertainties, either mechanism could account for the lack
of observed polarization.  If beam depolarization were the cause, it
would imply a significant small scale magnetic field with structure
at $l>10,000$.  Others have reported depth depolarization at similar
RM$\lambda^2$, so depth depolarization may play a role here as well.
When compared to the sensitivities required for EoR observations, the
upper bounds we report are still two orders of magnitude larger than
the sensitivities required.  The measured polarization leakage at GMRT
is less than 10\%,  so a modest leakage correction, accurate to 10\%,
should make EoR observations feasible.  The low polarized emission we
report is promising for the search for Reionization.

\section*{Acknowledgements}

We acknowledge helpful discussions with Ethan Vishniac, Michel
Brentjens, Marc Antoine Miville-Deschenes and Judd Bowman.

We acknowledge financial support by CIfAR, NSERC and NSF.  The GMRT is
operated by the National Centre for Radio Astrophysics (NCRA).  We thank
the NCRA for their extended support in this project.

\bibliography{polarized} 

\bibliographystyle{mn2e}

\appendix

\section{Power spectrum conventions}
\label{app:convention}

We report our final results in this paper in terms of (limits on) polarized power spectra.  This is appropriate because the theoretical predictions for 
21 cm fluctuations from the reionization epoch are usually presented as power spectra, either 2-dimensional ($C_l$) or 3-dimensional [$P(k)$].  An 
additional advantage is that unlike many other statistical measures, power spectra can be measured unambiguously over some range of $l$ (or $k$) from 
data that are missing zero-length baselines or contain significant instrument noise.  The purpose of this appendix is to summarize the definition of 
the power spectrum for both total intensity and polarization, and show how they can be derived from visibilities.

\subsection{2D power spectra: total intensity}

We first consider the 2-dimensional case, e.g. where we consider diffuse emission at only one frequency.  The intensity $I({\bmath n})$ is then a 
function of angular position ${\bmath n}$.  It can be Fourier-decomposed as
\begin{equation}
I({\bmath n}) - \bar I = \int \frac{{\rm d}^2{\bmath l}}{(2\pi)^2} \tilde I({\bmath l}) {\rm e}^{{\rm i}{\bmath l}\cdot{\bmath n}}.
\label{eq:fourier}
\end{equation}
Here ${\bmath l}$ is the Fourier wavenumber traditionally used in cosmology; it is related to ${\bmath u}=(u,v)$
(which has units of waves per radian) by ${\bmath l} = 2\pi {\bmath u}$.  Finally $\bar I$ is the mean intensity of the field.

The power spectrum $C_l$ of a diffuse field is a statistical measure defined by
\begin{equation}
\langle \tilde I^\ast({\bmath l}) \tilde I({\bmath l}') \rangle = (2\pi)^2 C_l \delta^{(2)}({\bmath l}-{\bmath l}'),
\label{eq:Cl}
\end{equation}
where $\delta^{(2)}$ is the 2-dimensional Dirac delta function.  It is related to the correlation function of the field by use of 
Eq.~(\ref{eq:fourier}),
\begin{equation}
\langle [I({\bmath n}) - \bar I][I({\bmath n}') - \bar I] \rangle
= \int \frac{{\rm d}^2{\bmath l}}{(2\pi)^2} C_l {\rm e}^{{\rm i}{\bmath l}\cdot({\bmath n}-{\bmath n}')}.
\end{equation}
In particular, the RMS intensity fluctuations in the map can be written as
\begin{equation}
\langle [I({\bmath n}) - \bar I]^2\rangle
= \int \frac{{\rm d}^2{\bmath l}}{(2\pi)^2} C_l 
= \int \frac{l^2C_l}{2\pi} \frac{{\rm d} l}l,
\label{eq:2d-int}
\end{equation}
where in the last equality we have integrated over the direction of ${\bmath l}$ and noted that the area element ${\rm d}^2{\bmath l}$ in the 
${\bmath l}$-plane is $2\pi l\,{\rm d} l$.  This integral is the reason why $l^2C_l/(2\pi)$ is a commonly used quantity: it describes the contribution 
to the intensity variance coming from a particular range of Fourier modes (specifically, per $\ln l$).

The power spectrum can also describe the RMS fluctuations of a sky map smoothed by a beam.  For example, convolving $I({\bmath n})$ with a Gaussian 
beam of FWHM $\theta_b$ to generate a smoothed map $I_{\rm sm}({\bmath n})$ is equivalent to multiplying its Fourier transform by the Fourier transform 
of a Gaussian:
\begin{equation}
\tilde I_{\rm sm}({\bmath l}) = {\rm e}^{-\theta_b^2 l^2/(16\ln2)}\tilde I({\bmath l}).
\end{equation}
The power spectrum of the smoothed map is then $[{\rm e}^{-\theta_b^2 l^2/(16\ln2)}]^2$ times the power spectrum of the original map, so the RMS 
fluctuation of the smoothed map is
\begin{equation}
\langle [I_{\rm sm}({\bmath n}) - \bar I]^2\rangle = \int \frac{l^2C_l}{2\pi}{\rm e}^{-\theta_b^2l^2/(8\ln2)} \frac{{\rm d} l}l.
\label{eq:2d-int-sm}
\end{equation}

We finally relate the power spectrum to the variance of visibilities observed on an interferometer.  We first consider the visibility $V({\bmath u})$ 
observed on an interferometer with baseline ${\bmath u}$ (separation transverse to line of sight in units of the wavelength $\lambda$), primary beam 
FWHM $\theta_b$.  For our angular coordinate ${\bmath n}$, we take the origin to be at the center of the primary beam.  A point source at position 
${\bmath n}$ with flux $F$ (in Jy) will give a visibility
\begin{equation}
V({\bmath u}) = {\rm e}^{-2\pi{\rm i}{\bmath u}\cdot{\bmath n}} {\rm e}^{-(4\ln 2)|{\bmath n}|^2/\theta_b^2} F,
\end{equation}
where the first factor gives the fringe phase and the second is the primary beam suppression.  In a diffuse field, the observed visibility will be the 
integral of the intensity over the field of view,
\begin{equation}
V({\bmath u}) = \int {\rm e}^{-2\pi{\rm i}{\bmath u}\cdot{\bmath n}} {\rm e}^{-(4\ln 2)|{\bmath n}|^2/\theta_b^2} I({\bmath n}){\rm d}^2{\bmath n}.
\end{equation}
Using Eq.~(\ref{eq:fourier}), and noting that the rapid modulation of the fringe phase implies that the mean intensity $\bar I$ gives no contribution, 
we find
\begin{eqnarray}
V({\bmath u}) \!\! &=& \!\! \int \frac{{\rm d}^2{\bmath l}}{(2\pi)^2} \tilde I({\bmath l})
  \int {\rm e}^{-2\pi{\rm i}{\bmath u}\cdot{\bmath n}} {\rm e}^{-(4\ln 2)|{\bmath n}|^2/\theta_b^2} {\rm e}^{{\rm i}{\bmath l}\cdot{\bmath n}}
  {\rm d}^2{\bmath n}
\nonumber \\
&=& \!\! \frac{\theta_b^2}{16\pi\ln2} \int {\rm d}^2{\bmath l}\, \tilde I({\bmath l})
   {\rm e}^{-\theta_b^2|{\bmath l}-2\pi{\bmath u}|^2/(16\ln 2)}.
\end{eqnarray}
The variance of this visibility can be found from Eq.~(\ref{eq:Cl}):
\begin{equation}
\langle |V({\bmath u})|^2 \rangle =
\left( \frac{\theta_b^2}{8\ln2}\right)^2 \int {\rm d}^2{\bmath l}\,C_l {\rm e}^{-\theta_b^2|{\bmath l}-2\pi{\bmath u}|^2/(8\ln 2)}.
\end{equation}
Thus the variance of the visibilities is directly related to the power spectrum at values of ${\bmath l}$ within $\sim\theta_b^{-1}$ of $2\pi{\bmath 
u}$.  In the practical case that the fringe spacing is much smaller than the primary beam ($\theta_b|{\bmath u}|\gg 1$) and the power spectrum varies 
slowly with scale, we may approximate $C_l$ by its value at $2\pi{\bmath u}$ and write
\begin{equation}
\langle |V({\bmath u})|^2 \rangle \approx \frac{\pi\theta_b^2}{8\ln2} C_{{\bmath l}=2\pi\bmath u}.
\label{eq:vis}
\end{equation}
This implies that the power spectrum can be determined in terms of the visibilities:
\begin{equation}
\frac{l^2}{2\pi}C_l|_{{\bmath l}=2\pi\bmath u} = \frac{16\ln2}{\theta_b^2}|{\bmath u}|^2 \langle |V({\bmath u})|^2 \rangle.
\label{eq:cvis}
\end{equation}

In our case, if the visibilities are calibrated in Jy, then the intensity $I({\bmath I})$ has units of Jy$\,$sr$^{-1}$ and the
power spectrum is determined in Jy$^{2}\,$sr$^{-2}$.  However we want the power spectrum in units of K$^2$, which means that the visibilities have to 
be converted to K$\,$sr units.  The conversion factor is
\begin{equation}
1\,{\rm Jy} = 0.00145\left(\frac{150\,\rm MHz}\nu\right)^2 {\rm K}\,{\rm sr}.
\end{equation}
Inserting this conversion factor into $V({\bmath u})$ in Eq.~(\ref{eq:cvis}) yields
\begin{equation}
\frac{l^2}{2\pi}C_{l\;{\rm K}^2}|_{{\bmath l}=2\pi\bmath u} = 2.33\times 10^{-5}
\frac{ |{\bmath u}|^2 }{ \theta_b^2 } \left(\frac{150\,\rm MHz}\nu\right)^4
\langle |V_{\rm Jy}({\bmath u})|^2 \rangle.
\label{eq:cvis2}
\end{equation}
If we refer the beam to the intrinsic FWHM of the GMRT primary beam, $3.3^\circ$ or 0.058 radians, this becomes
\begin{equation}
\frac{l^2}{2\pi}C_{l\;{\rm K}^2}|_{{\bmath l}=2\pi\bmath u} =
\left(\frac{|{\bmath u}|}{11.9} \frac{3.3^\circ}{\theta_b}\right)^2
\left(\frac{150\,\rm MHz}\nu\right)^4
\langle |V_{\rm Jy}({\bmath u})|^2 \rangle.
\end{equation}

Unlike e.g. the microwave background or extragalactic large scale structure, Galactic foregrounds are not statistically isotropic.  Nevertheless, one 
may still consider the power spectrum in a particular region of sky using Eq.~(\ref{eq:cvis2}).

\subsection{2D power spectra: polarization}

The same Fourier decomposition used for the total intensity can also be applied to linear polarization.
One may decompose the linear polarization in analogy to Eq.~(\ref{eq:fourier}) as
\begin{equation}
Q({\bmath n}) = \int \frac{{\rm d}^2{\bmath l}}{(2\pi)^2} \tilde Q({\bmath l}) {\rm e}^{{\rm i}{\bmath l}\cdot{\bmath n}},
\end{equation}
and similarly for $U$.

In studies of the microwave background, it is common to decompose the polarization into so-called $E$-mode and $B$-mode components, in which one 
rotates the reference axes of the Stokes parameters to be aligned with the Fourier wavevector ${\bmath l}$:
\begin{eqnarray}
\tilde Q({\bmath l}) &=& \cos (2\alpha_{\bmath l}) \tilde E({\bmath l}) - \sin (2\alpha_{\bmath l}) \tilde B({\bmath l}),
\nonumber \\
\tilde U({\bmath l}) &=& \sin (2\alpha_{\bmath l}) \tilde E({\bmath l}) + \cos (2\alpha_{\bmath l}) \tilde B({\bmath l}),
\end{eqnarray}
where $\alpha_{\bmath l}=\tan^{-1}(l_y/l_x)$ \citep{1996astro.ph..9149S,
1997ApJ...482....6S, 1997PhRvD..55.1830Z, 1997PhRvD..55.7368K}.\footnote{Some early papers denoted the rotated polarizations by $G$ and $C$, or 
$\oplus$ and $\otimes$, instead of $E$ and $B$, or used different normalizations.  Also note that this decomposition of polarization also applies to 
other spin 2 fields such as gravitational lensing shear.}  One can then define power spectra $C_l^E$ and 
$C_l^B$ in analogy to Eq.~(\ref{eq:Cl}).  This is useful 
because in the case of the microwave background the source of polarization is Thomson scattering, and hence the polarization direction is correlated 
with spatial morphology, leading to different power spectra in the $E$ and $B$ modes.  For low-frequency Galactic emission, however, Faraday rotation 
makes it unlikely that the observed polarization angles would have anything to do with the spatial morphology.  Therefore we instead work in terms of 
the sum of the power spectra, $C_l^P\equiv C_l^E+C_l^B$, which incorporates both Stokes parameters.  Then we have
\begin{equation}
\langle \tilde Q^\ast({\bmath l})\tilde Q({\bmath l}') + \tilde U^\ast({\bmath l})\tilde U({\bmath l}') \rangle =
(2\pi)^2C_l^P\delta^{(2)}({\bmath l}-{\bmath l}'),
\end{equation}
and the variance of the polarization, analogous to Eq.~(\ref{eq:2d-int}), is
\begin{equation}
\langle Q^2({\bmath n}) + U^2({\bmath n}) \rangle = \int \frac{l^2C_l^P}{2\pi} \frac{{\rm d}l}l.
\end{equation}
Thus $l^2C_l^P/(2\pi)$ can be viewed as the contribution to the polarization variance per $\ln l$.

The same methodology for estimating the intensity power spectrum from the variance of the visibilities can be applied to the polarization.  We find:
\begin{equation}
\frac{l^2}{2\pi}C^P_{l\;{\rm K}^2}|_{{\bmath l}=2\pi\bmath u} =
\left(\frac{|{\bmath u}|}{11.9} \frac{3.3^\circ}{\theta_b}\right)^2\!
\left(\frac{150\,\rm MHz}\nu\right)^4\!
\langle |V_{LR,\rm Jy}({\bmath u})|^2 \rangle.
\end{equation}

\subsection{3D power spectra: total intensity}

The predicted 21 cm signal is 3-dimensional in the sense of having structure in both the angular and radial (frequency) directions.  The statistic most commonly computed by theorists is power spectrum, 
$P(k,\mu)$, and the power per logarithmic range in wavenumber, $k^3P(k,\mu)/(2\pi^2)$ \citep{2004MNRAS.352..142B, 2005ApJ...624L..65B}.
To define this statistic, we begin by decomposing the intensity into its Fourier modes in analogy to Eq.~(\ref{eq:fourier3}):
\begin{equation}
I({\bmath r}) - \bar I = \int \frac{{\rm d}^3{\bmath k}}{(2\pi)^2} \tilde I({\bmath k}) {\rm e}^{{\rm i}{\bmath k}\cdot{\bmath r}};
\label{eq:fourier3}
\end{equation}
here ${\bmath r}$ is a 3-dimensional comoving position and ${\bmath k}$ is the wave vector.  Then a power spectrum is defined via:
\begin{equation}
\langle \tilde I^\ast({\bmath k}) \tilde I({\bmath k}') \rangle = (2\pi)^3 P({\bmath k}) \delta^{(3)}({\bmath k}-{\bmath k}').
\label{eq:Pk}
\end{equation}

Even in the case of a cosmological signal, the power spectrum may depend on the direction of ${\bmath k}$ relative to the line-of-sight vector which 
we take to be $\hat{\bmath e}_3$.  This is because the 
observed frequency of emission depends not just on the location of the emitting hydrogen gas, but also on its radial velocity.\footnote{An analogous 
situation occurs in galaxy redshift surveys \citep{1987MNRAS.227....1K}.}  We therefore introduce
\begin{equation}
\mu = \cos\theta_{\bmath k} = \frac{k_\parallel}{k} = \frac{{\bmath k}\cdot\hat{\bmath e}_3}{|{\bmath k}|},
\end{equation}
which is between $-1$ and $+1$.  Then the total variance of fluctuations is
\begin{equation}
\langle [I({\bmath r}) - \bar I]^2\rangle
= \int \frac{{\rm d}^3{\bmath k}}{(2\pi)^3} P({\bmath k})
= \int_0^\infty \frac{{\rm d}k}k \int_{-1}^1 \frac{{\rm d}\mu}2 \frac{k^3P(k,\mu)}{2\pi^2},
\label{eq:3d-int}
\end{equation}
where we have used that the volume element ${\rm d}^3{\bmath k}$ is $2\pi k^2\,{\rm d}\mu\,{\rm d}k$ after integrating out the azimuthal angle.  For 
an isotropic power spectrum, $P(k,\mu)=P(k)$, we 
can thus interpret $k^3P(k)/(2\pi^2)$ as the variance in intensity per logarithmic range in wavenumber.  Often one denotes $\Delta^2(k,\mu)=
k^3P(k,\mu)/(2\pi^2)$.  
The power spectrum is also sometimes written in components transverse to and radial along the line of sight, 
$P(k_\perp,k_\parallel)$.

In order to connect $P(k,\mu)$ to observations, we need to find the conversion from the comoving coordinates ${\bmath r}$ to the observables, angular 
position ${\bmath n}$ and frequency $\nu$.  In the direction transverse to the line of sight, the relation is ${\bmath r}_\perp = D{\bmath n}$, where 
$D$ is the comoving angular diameter distance to redshift $z=\lambda/(21{\rm cm})-1$.  In the direction along the line of sight, the relation is more 
complicated:
\begin{equation}
\frac{{\rm d}{\bmath r}_\parallel}{{\rm d}\nu} = \frac{{\rm d}{\bmath r}_\parallel/{\rm d}z}{{\rm d}\nu/{\rm d}z}
= \frac{(cH)^{-1}}{-\nu/(1+z)} = -\frac{1+z}{cH\nu},
\end{equation}
where $c$ is the speed of light, $H$ is the Hubble constant at redshift $z$, and $\nu$ is the observed frequency.  We can then define the radial 
conversion factor $C\equiv 
(1+z)/(cH\nu)$.  It follows that the comoving coordinates are related to observed coordinates via $({\bmath r}_\perp, r_\parallel) \leftrightarrow 
[D{\bmath n},-C(\nu-\nu_0)]$, where $\nu_0$ is a reference frequency (the zero point of $r_\parallel$ is arbitrary).

We can now construct the 2-dimensional power spectrum, which can be measured via Eq.~(\ref{eq:cvis}), in terms of the 3-dimensional power spectrum.  We 
construct a 2-dimensional image $I_{\rm 2D}({\bmath n})$ with bandwidth $\Delta\nu$,
\begin{eqnarray}
I_{\rm 2D}({\bmath n}) &=& \frac1{\Delta\nu} \int_{\nu_0-\Delta\nu/2}^{\nu_0+\Delta\nu/2} I({\bmath n},\nu)\,{\rm d}\nu
\nonumber \\
&=& \frac1{C\Delta\nu} \int_{-C\Delta\nu/2}^{C\Delta\nu/2} I(D{\bmath n},r_\parallel)\,{\rm d}r_\parallel.
\end{eqnarray}
Using Eq.~(\ref{eq:fourier}), we can find the Fourier transform of the 2-dimensional map.  For ${\bmath l}\neq 0$,
\begin{eqnarray}
\tilde I_{\rm 2D}({\bmath l}) &=& \int {\rm d}^2{\bmath n}\; I({\bmath n}) {\rm e}^{-{\rm i}{\bmath l}\cdot{\bmath n}}
\nonumber \\
&=& \frac1{C\Delta\nu} \int {\rm d}^2{\bmath n} \int_{-C\Delta\nu/2}^{C\Delta\nu/2}{\rm d}r_\parallel\; I(D{\bmath n},r_\parallel)
  {\rm e}^{-{\rm i}{\bmath l}\cdot{\bmath n}}
\nonumber \\
&=& \frac1{C\Delta\nu} \int {\rm d}^2{\bmath n} \int_{-C\Delta\nu/2}^{C\Delta\nu/2}{\rm d}r_\parallel \int \frac{ {\rm d}^2{\bmath k}_\perp\,
{\rm d}k_\parallel}{(2\pi)^3} \tilde I({\bmath k})
  \nonumber \\ &&\times {\rm e}^{{\rm i}({\bmath k}_\perp\cdot D{\bmath n}+k_\parallel r_\parallel)}
  {\rm e}^{-{\rm i}{\bmath l}\cdot{\bmath n}}
\nonumber \\
&=& \frac1{C\Delta\nu} \int_{-C\Delta\nu/2}^{C\Delta\nu/2}{\rm d}r_\parallel \int \frac{ {\rm d}^2{\bmath k}_\perp\,
{\rm d}k_\parallel}{(2\pi)^3} \tilde I({\bmath k})
  \nonumber \\ && \times (2\pi)^2\delta^{(2)}(D{\bmath k}_\perp-{\bmath l}) {\rm e}^{{\rm i}k_\parallel r_\parallel}
\nonumber \\
&=& \frac1{D^2} \int \frac{{\rm d}k_\parallel}{2\pi} \tilde I\left(\frac{\bmath l}D,k_\parallel\right)
 \sinc\frac{C\Delta\nu\,k_\parallel}2,
\end{eqnarray}
where $\sinc\,x\equiv\sin x/x$.  In the last equality, the $\,{\rm d}^2{\bmath k}_\perp$ integral is trivial due to the $\delta$-function, and the 
$r_\parallel$ integral gives the sinc function.  By taking the variance we extract the 2-dimensional power spectrum,
\begin{equation}
C_l^{\rm 2D} = \frac1{D^2} \int \frac{{\rm d}k_\parallel}{2\pi} P\left(\frac{\bmath l}D,k_\parallel\right)
\sinc^2\frac{C\Delta\nu\,k_\parallel}2.
\label{eq:project}
\end{equation}
Note that the amount of power in the 2-dimensional map depends on the bandwidth $\Delta\nu$.  This is a direct consequence of the 3-dimensional nature 
of the EoR signal: fluctuations with $|k_\parallel|>(C\Delta\nu)^{-1}$ are cancelled by the combination of frequencies.

It is possible to measure the full 3-dimensional power spectrum by stacking maps at several neighboring frequencies with a phase shift to avoid 
cancellation of frequencies, i.e.
\begin{equation}
I_{\rm 2D}({\bmath n},\tau) = \frac1{\Delta\nu} \int_{\nu_0-\Delta\nu/2}^{\nu_0+\Delta\nu/2} {\rm e}^{2\pi{\rm i}\tau\nu}
 I({\bmath n},\nu)\,{\rm d}\nu,
\label{eq:2d-tau}
\end{equation}
where $\tau$ is the phase shift in units of cycles per MHz.  This is now a complex map but this causes no difficulty.  The power spectrum of $I_{\rm 
2D}({\bmath n},\tau)$ can be obtained by the same methodology as Eq.~(\ref{eq:project}), yielding
\begin{equation}
C_l^{\rm 2D}(\tau) = \frac1{D^2} \int \frac{{\rm d}k_\parallel}{2\pi} P\left(\frac{\bmath l}D,k_\parallel\right)
\sinc^2\frac{\Delta\nu(Ck_\parallel-2\pi\tau)}2.
\label{eq:project-tau}
\end{equation}
Thus this stacked map is sensitive to fluctuations with radial structure and wavenumbers in the range $k_\parallel\approx 2\pi\tau/C$, witihin a width
$\delta k_\parallel \sim (C\Delta\nu)^{-1}$.  If the power spectrum is slowly varying as a function of $k_\parallel$ over this range, one may 
use the fact that $\int \sinc^2 x\,dx = \pi$ and approximate
\begin{equation}
C_l^{\rm 2D}(\tau) \approx \frac{1}{D^2C\Delta\nu} P\left(\frac{\bmath l}D,\frac{2\pi\tau}C\right).
\label{eq:project3}
\end{equation}
By combining with Eq.~(\ref{eq:cvis2}), this allows one to estimate the 3-dimensional power spectrum from frequency-stacked 2-dimensional visibilities.

\subsection{3D power spectra: polarization}
\label{ssa:3dp}

The machinery used to describe intensity fluctuations can also be used to describe polarization fluctuations.  Polarized emission from EoR is expected 
to be intrinsically very small and if present would be strongly affected by Galactic Faraday rotation (although see \citealt{2005ApJ...635....1B}).  Polarized 
emission from the Galaxy is a potentially severe foreground, and we characterize it by describing its power spectrum using the same conventions used 
for the EoR signal.  To do this, we decompose the polarized flux density into Fourier modes just as was done for the intensity:
\begin{equation}
Q({\bmath r}) = \int \frac{{\rm d}^3{\bmath k}}{(2\pi)^2} \tilde Q({\bmath k}) {\rm e}^{{\rm i}{\bmath k}\cdot{\bmath r}},
\label{eq:fourier3p}
\end{equation}
and similarly for $U$.  One may define a polarized power spectrum via
\begin{equation}
\langle \tilde Q^\ast({\bmath k}) \tilde U({\bmath k}') \rangle
= (2\pi)^3 P_{QU}({\bmath k}) \delta^{(3)}({\bmath k}-{\bmath k}'),
\end{equation}
and similarly for $P_{QQ}$, $P_{UQ}$, and $P_{UU}$.  Since $Q$ and $U$ are real it follows that $P_{QQ}$ and $P_{UU}$ are real and non-negative
and satisfy $P_{QQ}({\bmath k}) = P_{QQ}(-{\bmath k})$ and $P_{UU}({\bmath k})=P_{UU}(-{\bmath k})$; also
$P_{QU}({\bmath k}) = P_{UQ}^\ast(-{\bmath k})$.
A decomposition of the polarization tensor into $E$ and $B$ modes is possible, just as in the two-dimensional case, but is probably not interesting if 
the radiation has propagated through a Faraday thick medium.  We therefore focus on measuring the power spectrum $P_P({\bmath k})=P_{QQ}+P_{UU}$:
\begin{equation}
\langle \tilde Q^\ast({\bmath k}) \tilde Q({\bmath k}') + \tilde U^\ast({\bmath k}) \tilde U({\bmath k}') \rangle
= (2\pi)^3 P_P({\bmath k}) \delta^{(3)}({\bmath k}-{\bmath k}').
\label{eq:pol3d}
\end{equation}

The 3-dimensional polarized power spectrum can be obtained by stacking in analogy to Eq.~(\ref{eq:2d-tau}).  We define stacked 2-dimensional
polarization maps,
\begin{equation}
P_{\rm 2D}({\bmath n},\tau) = \frac1{\Delta\nu} \int_{\nu_0-\Delta\nu/2}^{\nu_0+\Delta\nu/2} {\rm e}^{2\pi{\rm i}\tau\nu}
 [Q({\bmath n},\nu)+{\rm i}U({\bmath n},\nu)]\,{\rm d}\nu.
\label{eq:2d-tau-p}
\end{equation}
Repeating the steps that lead to Eq.~(\ref{eq:project3}) gives, in the polarization case, that the 2-dimensional power spectra of
the complex maps $P_{\rm 2D}({\bmath n},\tau)$ are
\begin{eqnarray}
C_l^{P, \rm 2D}(\tau)\!\! &\approx& \!\! \frac{1}{D^2C\Delta\nu}
\nonumber \\ && \!\! \times
\left[
P_P\left(\frac{\bmath l}D,\frac{2\pi\tau}C\right)
-2\Im P_{QU}\left(\frac{\bmath l}D,\frac{2\pi\tau}C\right)
\right],
\nonumber \\ &&
\label{eq:project3p}
\end{eqnarray}
where $\Im$ denotes the imaginary part.  Note that for a signal that respects statistical parity invariance, the second term here must vanish, but this 
may not be true for the Galactic emission in a particular patch 
of sky.  This term may be removed by averaging signals from positive and negative $\tau$.

In the case of the cosmological signal in the unpolarized intensity, we identified $k_\parallel$ with the line-of-sight wavenumber of a mode in large scale structure.  In the case of polarized Galactic 
synchrotron radiation, the rapid variation of the signal with frequency is instead due to Faraday rotation.  In fact, Eq.~(\ref{eq:2d-tau-p}) is (at least over a limited range of frequency) an 
implementation of RM synthesis, with the exponent ${\rm e}^{-2{\rm i}\phi \lambda^2}$ replaced by ${\rm e}^{2\pi{\rm i}\tau\nu}$.  That is, it is the RM synthesis map at Faraday depth
\begin{equation}
\phi = -\pi\frac{{\rm d}\nu}{{\rm d}(\lambda^2)}\tau = \frac{\pi\nu}{2\lambda^2}\tau.
\end{equation}
Using the conversions from $\tau$ to $k_\parallel$ ($k_\parallel=2\pi\tau/C$), this becomes
\begin{equation}
\phi = \frac{C\nu}{4\lambda^2}k_\parallel = \frac{1+z}{4cH\lambda^2}k_\parallel.
\label{eq:phi-k}
\end{equation}

We thus conclude that the mathematical procedure to extract the EoR power spectrum from a foreground-cleaned data cube is (in principle!) similar to 
that of computing a 2-dimensional power spectrum on RM synthesis maps, except that it is applied to the real total intensity $I$ [via 
Eq.~(\ref{eq:2d-tau})] rather than complex polarization $Q+{\rm i}U$.  In particular, a simple type of polarization-to-intensity leakage such as 
\begin{equation}
I_{\rm obs}(u,v,\nu)=I(u,v,\nu)+\gamma_Q Q(u,v,\nu) + \gamma_U U(u,v,\nu)
\end{equation}
will leak polarized structure at Faraday depth $\phi$ into the intensity map at wavenumber 
$k_\parallel$ given by Eq.~(\ref{eq:phi-k}) if $(\gamma_Q,\gamma_U)$ is frequency-independent (or varies slowly with frequency), since it results in an 
apparent 
Stokes $I$ signal with a sinusoidal variation in $\nu$.  In this case, the observed power spectrum would be
\begin{equation}
P_{\rm obs}({\bmath k}) = P({\bmath k}) + \gamma_Q^2P_{QQ}({\bmath k})
  + \gamma_U^2P_{UU}({\bmath k}) + 2\gamma_Q\gamma_U \Re P_{QU}({\bmath k}).
\end{equation}
More complex types of frequency-dependent polarization leakage can occur in real EoR observations, especially if observations at the same $(u,v)$ but 
different frequencies are taken non-simultaneously [as is necessary at GMRT, since a given point in the $(u,v)$ plane corresponds to a different 
physical baseline length at a different frequency].  One example would be polarized sidelobes that rotate relative to the sky due to the alt-azimuth 
mount.  Another would be changes in ionospheric Faraday rotation between the epochs of observation.  Finally the beam squint will vary across the 
bandpass.
These frequency-dependent leakages are a concern because polarization at any Faraday depth can be injected into the Stokes $I$ map at wavenumbers that
do not obey Eq.~(\ref{eq:phi-k}).

\end{document}